# A Review of Accelerated Test Models

**Luis A. Escobar and William Q. Meeker**

*Abstract.* Engineers in the manufacturing industries have used accelerated test (AT) experiments for many decades. The purpose of AT experiments is to acquire reliability information quickly. Test units of a material, component, subsystem or entire systems are subjected to higher-than-usual levels of one or more accelerating variables such as temperature or stress. Then the AT results are used to predict life of the units at use conditions. The extrapolation is typically justified (correctly or incorrectly) on the basis of physically motivated models or a combination of empirical model fitting with a sufficient amount of previous experience in testing similar units. The need to extrapolate in both time and the accelerating variables generally necessitates the use of fully parametric models. Statisticians have made important contributions in the development of appropriate stochastic models for AT data [typically a distribution for the response and regression relationships between the parameters of this distribution and the accelerating variable(s)], statistical methods for AT planning (choice of accelerating variable levels and allocation of available test units to those levels) and methods of estimation of suitable reliability metrics. This paper provides a review of many of the AT models that have been used successfully in this area.

## 1. INTRODUCTION

### 1.1 Motivation

Today's manufacturers face strong pressure to develop new, higher-technology products in record time,

*Luis A. Escobar is Professor, Department of Experimental Statistics, Louisiana State University, Baton Rouge, Louisiana 70803, USA e-mail: luis@lsu.edu. William Q. Meeker is Distinguished Professor, Department of Statistics, Iowa State University, Ames, Iowa 50011, USA e-mail: wqmeeker@iastate.edu.*

while improving productivity, product field reliability and overall quality. This has motivated the development of methods like concurrent engineering and encouraged wider use of designed experiments for product and process improvement. The requirements for higher reliability have increased the need for more *up-front* testing of materials, components and systems. This is in line with the modern quality philosophy for producing high-reliability products: achieve high reliability by improving the design and manufacturing processes; move away from reliance on inspection (or screening) to achieve high reliability, as described in Meeker and Hamada (1995) and Meeker and Escobar (2004).

Estimating the failure-time distribution or long-term performance of components of *high-reliability* products is particularly difficult. Most modern products are designed to operate without failure for years, decades or longer. Thus few units will fail or degrade appreciably in a test of practical length at normal use conditions. For example, the design and construction of a communications satellite may allow only eight months to test components that are expected to be in service for 10 or 15 years. For such applications, Accelerated Tests (ATs) are used in manufacturing industries to assess or demonstrate component and subsystem reliability, to certify components, to detect failure modes so that they can be corrected, to compare different manufacturers, and so forth. ATs have become increasingly important because of rapidly changing technologies, more complicated products with more components, higher customer expectations for better reliability and the need for rapid product development. There are difficult practical and statistical issues involved in accelerating the life of a complicated product that can fail in different ways. Generally, information from tests at high levels of one or more accelerating variables (e.g., use rate, temperature, voltage or pressure) is extrapolated, through a physically reasonable statistical model, to obtain estimates of life or long-term performance at lower, normal levels of the accelerating variable(s).

Statisticians in manufacturing industries are often asked to become involved in planning or analyzing data from accelerated tests. At first glance, the statistics of accelerated testing appears to involve little more than regression analysis, perhaps with a few complicating factors, such as censored data. The very nature of ATs, however, always requires extrapolation in the accelerating variable(s) and often requires extrapolation in time. This implies critical importance of model choice. Relying on the common statistical practice of fitting curves to data can result in an inadequate model or even nonsense results. Statisticians working on AT programs need to be aware of general principles of AT modeling and current best practices.

The purpose of this review paper is to outline some of the basic ideas behind accelerated testing and especially to review currently used AT modeling practice and to describe the most commonly used AT models. In our concluding remarks we make explicit suggestions about the potential contributions that statisticians should be making to the development of AT models and methods. We illustrate the use of the different models with a series of examples from the literature and our own experiences.

## 1.2 Quantitative versus Qualitative Accelerated Tests

Within the reliability discipline, the term "accelerated test" is used to describe two completely different kinds of useful, important tests that have completely different purposes. To distinguish between these, the terms "quantitative accelerated tests" (QuanAT) and "qualitative accelerated tests" (QualAT) are sometimes used.

A QuanAT tests units at combinations of higher-than-usual levels of certain accelerating variables. The purpose of a QuanAT is to obtain information about the failure-time distribution or degradation distribution at specified "use" levels of these variables. Generally failure modes of interest are known ahead of time, and there is some knowledge available that describes the relationship between the failure mechanism and the accelerating variables (either based on physical/chemical theory or large amounts of previous experience with similar tests) that can be used to identify a model that can be used to justify the extrapolation. In this paper, we describe models for QuanATs.

A QualAT tests units at higher-than-usual combinations of variables like temperature cycling and vibration. Specific names of such tests include HALT (for highly accelerated life tests), STRIFE (stress-life) and EST (environmental stress testing). The purpose of such tests is to identify product weaknesses caused by flaws in the product's design or manufacturing process. Nelson (1990, pages 37–39) describes such tests as "elephant tests" and outlines some important issues related to QualATs.

When there is a failure in a QualAT it is necessary to find and carefully study the failure's root cause and assess whether the failure mode could occur in actual use or not. Knowledge and physical/chemical modeling of the particular failure mode is useful for helping to make this assessment. When it is determined that a failure could occur in actual use, it is necessary to change the product design or manufacturing process to eliminate that cause of failure. Nelson (1990, page 38) describes an example where a costly effort was made to remove a high-stress-induced failure mode that never would have occurred in actual use.

Because the results of a QualAT are used to make changes on the product design or manufacturing process, it is difficult, or at the very least, very risky to use the test data to predict what will happen in normal use. Thus QualATs are generally thought of as being nonstatistical.

## 1.3 Overview

The rest of this paper is organized as follows. Section 2 describes the basic physical and practical ideas behind the use of ATs and the characteristics of various kinds of AT data. Section 3 describes the concept of a time-transformation model as an accelerated failure-time model, describes some commonly used special cases and also presents several nonstandard special cases that are important in practice. Section 4 describes acceleration models that are used when product use rate is increased to get information quickly. Sections 5 and 6 explain and illustrate the use of temperature and humidity, respectively, to accelerate failure mechanisms. Section 7 describes some of the special characteristics of ATs for photodegradation. Section 8 explains and illustrates the use of increased voltage (or voltage stress) in ATs with the commonly used inverse power relationship. This section also describes how a more general relationship, based on the Box–Cox transformation, can be used in sensitivity analyses that help engineers to make decisions. Section 9 describes examples in which combinations of two or more accelerating variables are used in an AT. Section 10 discusses some practical concerns and general guidelines for conducting and interpreting ATs. Section 11 describes areas of future research in the development of accelerated test models and the role that statisticians will have in these developments.

## 2. BASIC IDEAS OF ACCELERATED TESTING

### 2.1 Different Types of Acceleration

The term "acceleration" has many different meanings within the field of reliability, but the term generally implies making "time" (on whatever scale is used to measure device or component life) go more quickly, so that reliability information can be obtained more rapidly.

### 2.2 Methods of Acceleration

There are different methods of accelerating a reliability test:

*Increase the use rate of the product.* This method is appropriate for products that are ordinarily not in continuous use. For example, the median life of a bearing for a certain washing machine agitator is 12 years, based on an assumed use rate of 8 loads per week. If the machine is tested at 112 loads per week (16 per day), the median life is reduced to roughly 10 months, under the assumption that the increased use rate does not change the cycles to failure distribution. Also, because it is not necessary to have all units fail in a life test, useful reliability information could be obtained in a matter of weeks instead of months.

*Increase the intensity of the exposure to radiation.* Various types of radiation can lead to material degradation and product failure. For example, organic materials (ranging from human skin to materials like epoxies and polyvinyl chloride or PVC) will degrade when exposed to ultraviolet (UV) radiation. Electrical insulation exposed to gamma rays in nuclear power plants will degrade more rapidly than similar insulation in similar environments without the radiation. Modeling and acceleration of degradation processes by increasing radiation intensity is commonly done in a manner that is similar to acceleration by increasing use rate.

*Increase the aging rate of the product.* Increasing the level of experimental variables like temperature or humidity can accelerate the chemical processes of certain failure mechanisms such as chemical degradation (resulting in eventual weakening and failure) of an adhesive mechanical bond or the growth of a conducting filament across an insulator (eventually causing a short circuit).

*Increase the level of stress (e.g., amplitude in temperature cycling, voltage, or pressure) under which test units operate.* A unit will fail when its *strength* drops below applied stress. Thus a unit at a high stress will generally fail more rapidly than it would have failed at low stress.

Combinations of these methods of acceleration are also employed. Variables like voltage and temperature cycling can both increase the rate of an electrochemical reaction (thus accelerating the aging rate) and increase stress relative to strength. In such situations, when the effect of an accelerating variable is complicated, there may not be enough physical knowledge to provide an adequate physical model for acceleration (and extrapolation). Empirical models may or may not be useful for extrapolation to use conditions.

### 2.3 Types of Responses

It is useful to distinguish among ATs on the basis of the nature of the response.

*Accelerated Binary Tests* (*ABTs*). The response in an ABT is binary. That is, whether the product has failed or not is the only reliability information obtained from each unit. See Meeker and Hahn (1977) for an example and references.

*Accelerated Life Tests* (*ALTs*). The response in an ALT is directly related to the lifetime of the product. Typically, ALT data are right-censored because the test is stopped before all units fail. In other cases, the ALT response is interval-censored because failures are discovered at particular inspection times. See Chapters 2–10 of Nelson (1990) for a comprehensive treatment of ALTs.

*Accelerated Repeated Measures Degradation Tests* (*ARMDTs*). In an ARMDT, one measures degradation on a sample of units at different points in time. In general, each unit provides several degradation measurements. The degradation response could be actual chemical or physical degradation or performance degradation (e.g., drop in power output). See Meeker and Escobar (1998, Chapters 13 and 21) for examples of ARMDT modeling and analysis.

*Accelerated Destructive Degradation Tests* (*ADDTs*). An ADDT is similar to an ARMDT, except that the measurements are destructive, so one can obtain only one observation per test unit. See Escobar, Meeker, Kugler and Kramer (2003) for a discussion of ADDT methodology and a detailed case study.

These different kinds of ATs can be closely related because they can involve the same underlying physical/chemical mechanisms for failure and models for acceleration. They are different, however, in that different kinds of statistical models and analyses are performed because of the differences in the kind of response that can be observed.

Many of the underlying physical model assumptions, concepts and practices are the same for ABTs, ALTs, ARMDTs and ADDTs. There are close relationships among models for ABT, ALT, ARMD and ADD data. Because of the different types of responses, however, the actual models fitted to the data and methods of analysis differ. In some cases, analysts use degradation-level data to define failure times. For example, turning ARMDT data into ALT data generally simplifies analysis but may sacrifice useful information. An important characteristic of all ATs is the need to extrapolate outside the range of available data: tests are done at accelerated conditions, but estimates are needed at use conditions.

Such extrapolation requires strong model assumptions.

# 3. STATISTICAL MODELS FOR ACCELERATION

This section discusses acceleration models and some physical considerations that lead to these models. For further information on these models, see Nelson (1990, Chapter 2) and Meeker and Escobar (1998, Chapter 18). Other useful references include Smith (1996), Section 7 of Tobias and Trindade (1995), Sections 2 and 9 of Jensen (1995) and Klinger, Nakada and Menendez (1990).

Interpretation of accelerated test data requires models that relate accelerating variables like temperature, voltage, pressure, size, etc. to time acceleration. For testing over some range of accelerating variables, one can fit a model to the data to describe the effect that the variables have on the failure-causing processes. The general idea is to test at high levels of the accelerating variable(s) to speed up failure processes and extrapolate to lower levels of the accelerating variable(s). For some situations, a physically reasonable statistical model may allow such extrapolation.

*Physical acceleration models.* For well-understood failure mechanisms, one may have a model based on physical/chemical theory that describes the failure-causing process over the range of the data and provides extrapolation to use conditions. The relationship between accelerating variables and the actual failure mechanism is usually extremely complicated. Often, however, one has a simple model that adequately describes the process. For example, failure may result from a complicated chemical process with many steps, but there may be one rate-limiting (or dominant) step and a good understanding of this part of the process may provide a model that is adequate for extrapolation.

*Empirical acceleration models.* When there is little understanding of the chemical or physical processes leading to failure, it may be impossible to develop a model based on physical/chemical theory. An empirical model may be the only alternative. An empirical model may provide an excellent fit to the available data, but provide nonsense extrapolations (e.g., the quadratic models used in Meeker and Escobar, 1998, Section 17.5). In some situations there may be extensive empirical experience with particular combinations of variables and failure mechanisms and this experience may provide the needed justification for extrapolation to use conditions.

In the rest of this section we will describe the general time-transformation model and some special acceleration models that have been useful in specific applications.

### 3.1 General Time-Transformation Functions

A time-transformation model maps time at one level of $\mathbf{x}$, say $\mathbf{x}_U$, to time at another level of $\mathbf{x}$. This can be expressed as $T(\mathbf{x}) = \Upsilon[T(\mathbf{x}_U), \mathbf{x}]$, where $\mathbf{x}_U$ denotes use conditions. To be a time transformation, the function $\Upsilon(t, \mathbf{x})$ must have the following properties:

- For any $\mathbf{x}$, $\Upsilon(0, \mathbf{x}) = 0$, as in Figure 1.
- $\Upsilon(t, \mathbf{x})$ is nonnegative, that is, $\Upsilon(t, \mathbf{x}) \geq 0$ for all $t$ and $\mathbf{x}$.
- For fixed $\mathbf{x}$, $\Upsilon(t, \mathbf{x})$ is monotone increasing in $t$.
- When evaluated at $\mathbf{x}_U$, the transformation is the identity transformation [i.e., $\Upsilon(t, \mathbf{x}_U) = t$ for all $t$].

A quantile of the distribution of $T(\mathbf{x})$ can be determined as a function of $\mathbf{x}$ and the corresponding quantile of the distribution of $T(\mathbf{x}_U)$. In particular, $t_p(\mathbf{x}) = \Upsilon[t_p(\mathbf{x}_U), \mathbf{x}]$ for $0 \leq p \leq 1$. As shown in Figure 1, a plot of $T(\mathbf{x}_U)$ versus $T(\mathbf{x})$ can imply a particular class of transformation functions. In particular,

- $T(\mathbf{x})$ entirely below the diagonal line implies acceleration.
- $T(\mathbf{x})$ entirely above the diagonal line implies deceleration.
- $T(\mathbf{x})$ can cross the diagonal, in which case the transformation is accelerating over some times and decelerating over other times. In this case the c.d.f.'s of $T(\mathbf{x})$ and $T(\mathbf{x}_U)$ cross. See Martin (1982) and LuValle, Welsher and Svoboda (1988) for further discussion of time-transformation models.

### 3.2 Scale-Accelerated Failure-Time Models (SAFTs)

A simple, commonly used model used to characterize the effect that explanatory variables $\mathbf{x} = (x_1, \ldots, x_k)'$ have on lifetime $T$ is the scale-accelerated failure-time (SAFT) model. The model is ubiquitous in the statistical literature where it is generally referred to as the "accelerated failure-time model." It is, however, a very special kind of accelerated failure-time model. Some of the explanatory variables in $\mathbf{x}$ are used for acceleration, but others may be of interest for other reasons (e.g., for product design optimization decisions). Under a SAFT model, lifetime at $\mathbf{x}$, $T(\mathbf{x})$, is scaled by a deterministic factor that might depend on $\mathbf{x}$ and unknown fixed parameters. More specifically, a model for the random variable $T(\mathbf{x})$ is SAFT if $T(\mathbf{x}) = T(\mathbf{x}_U)/\mathcal{AF}(\mathbf{x})$, where the *acceleration factor* $\mathcal{AF}(\mathbf{x})$ is a positive function of $\mathbf{x}$ satisfying $\mathcal{AF}(\mathbf{x}_U) = 1$. Lifetime is accelerated (decelerated) when $\mathcal{AF}(\mathbf{x}) > 1$ [$\mathcal{AF}(\mathbf{x}) < 1$]. In terms of distribution quantiles,

$$t_p(\mathbf{x}) = \frac{t_p(\mathbf{x}_U)}{\mathcal{AF}(\mathbf{x})}. \tag{1}$$

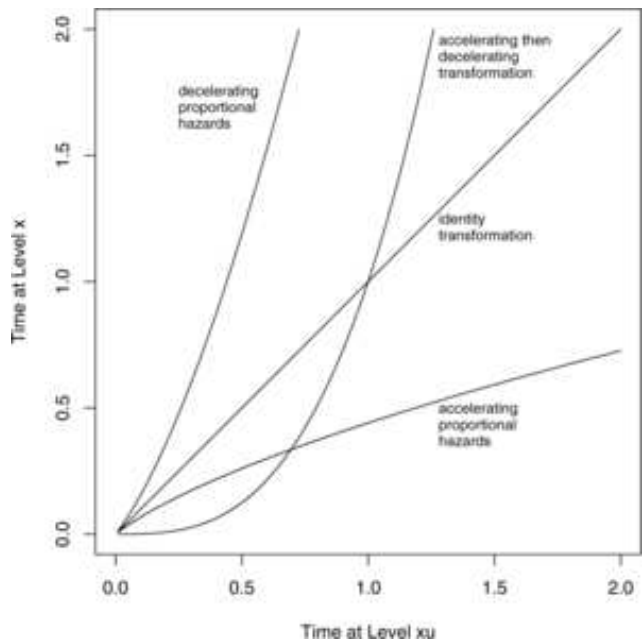

FIG. 1. *General failure-time transformation with $x_u < x$.*

Some special cases of these important SAFT models are discussed in the following sections.

Observe that under a SAFT model, the probability that failure at conditions $\mathbf{x}$ occurs at or before time $t$ can be written as $\Pr[T(\mathbf{x}) \leq t] = \Pr[T(\mathbf{x}_U) \leq \mathcal{AF}(\mathbf{x}) \times t]$. It is common practice (but certainly not necessary) to assume that lifetime $T(\mathbf{x})$ has a log-location-scale distribution, with parameters $(\mu, \sigma)$, such as a lognormal distribution in which $\mu$ is a function of the accelerating variable(s) and $\sigma$ is constant (i.e., does not depend on $\mathbf{x}$). In this case,

$$F(t; \mathbf{x}_U) = \Pr[T(\mathbf{x}_U) \leq t] = \Phi\left[\frac{\log(t) - \mu_U}{\sigma}\right],$$

where $\Phi$ denotes a standard cumulative distribution function (e.g., standard normal) and $\mu_U$ is the location parameter for the distribution of $\log[T(\mathbf{x}_U)]$.

Thus,

$$F(t; \mathbf{x}) = \Pr[T(\mathbf{x}) \leq t]$$
$$= \Phi\left(\frac{\log(t) - \{\mu_U - \log[\mathcal{AF}(\mathbf{x})]\}}{\sigma}\right).$$

Note that $T(\mathbf{x})$ also has a log-location-scale distribution with location parameter $\mu = \mu_U - \log[\mathcal{AF}(\mathbf{x})]$ and a scale parameter $\sigma$ that does not depend on $\mathbf{x}$.

### 3.3 The Proportional Hazard Regression Model

For a continuous cdf $F(t; \mathbf{x}_U)$ and $\Psi(\mathbf{x}) > 0$ define the time transformation

$$T(\mathbf{x}) = F^{-1}(1 - \{1 - F[T(\mathbf{x}_U); \mathbf{x}_U]\}^{1/\Psi(\mathbf{x})}; \mathbf{x}_U).$$

It can be shown that $T(\mathbf{x})$ and $T(\mathbf{x}_U)$ have the proportional hazard (PH) relationship

$$(2) \qquad h(t; \mathbf{x}) = \Psi(\mathbf{x})h(t; \mathbf{x}_U).$$

This time-transformation function is illustrated in Figure 1. In this example, the amount of acceleration (or deceleration), $T(\mathbf{x}_U)/T(\mathbf{x})$, depends on the position in time and the model is not a SAFT. If $F(t; \mathbf{x}_U)$ has a Weibull distribution with scale parameter $\eta_U$ and shape parameter $\beta_U$, then $T(\mathbf{x}) = T(\mathbf{x}_U)/\mathcal{AF}(\mathbf{x})$, where $\mathcal{AF}(\mathbf{x}) = [\Psi(\mathbf{x})]^{1/\beta_U}$. This implies that this particular PH regression model is also a SAFT regression model. It can be shown that the Weibull distribution is the only distribution in which both (1) and (2) hold. Lawless (1986) illustrates this result nicely.

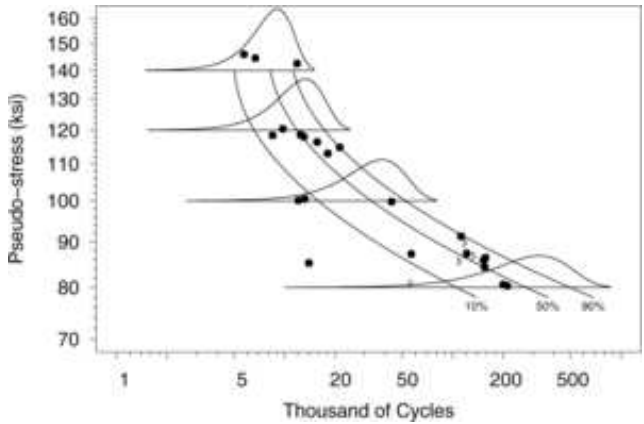

FIG. 2. *Superalloy fatigue data with fitted log-quadratic Weibull regression model with nonconstant $\sigma$. Censored observations are indicated by ▷. The response, cycles, is shown on the x-axis.*

### 3.4 Another Non-SAFT Example: The Nonconstant $\sigma$ Regression Model

This section describes acceleration models with nonconstant $\sigma$. In some lifetime applications, it is useful to consider log-location-scale models in which both $\mu$ and $\sigma$ depend on explanatory variables. The log-quantile function for this model is

$$\log[t_p(\mathbf{x})] = \mu(\mathbf{x}) + \Phi^{-1}(p)\sigma(\mathbf{x}).$$

Thus

$$\frac{t_p(\mathbf{x}_U)}{t_p(\mathbf{x})} = \exp\{\mu(\mathbf{x}_U - \mu(\mathbf{x}))$$
$$+ \Phi^{-1}(p)[\sigma(\mathbf{x}_U) - \sigma(\mathbf{x})]\}.$$

Because $t_p(\mathbf{x}_U)/t_p(\mathbf{x})$ depends on $p$, this model is not a SAFT model.

EXAMPLE 1 (*Weibull log-quadratic regression model with nonconstant $\sigma$ for the superalloy fatigue data*). Meeker and Escobar (1998, Section 17.5) analyze superalloy fatigue data using a Weibull model in which $\mu = \beta_0^{[\mu]} + \beta_1^{[\mu]}x + \beta_2^{[\mu]}x^2$ and $\log(\sigma) = \beta_0^{[\sigma]} + \beta_1^{[\sigma]}x$ (see Nelson, 1984 and 1990, for a similar analysis using a lognormal distribution). Figure 2 shows the log-quadratic Weibull regression model with nonconstant $\sigma$ fit to the superalloy fatigue data.

Meeker and Escobar (1998) indicate that the evidence for nonconstant $\sigma$ in the data is not strong. But having $\sigma$ decrease with stress or strain is typical in fatigue data and this is what the data points plotted in Figure 2 show. Thus, it is reasonable to use a model with decreasing $\sigma$ in this case, even in the absence of "statistical significance," especially because assuming a constant sigma could lead to anti-conservative estimates of life at lower levels of stress.

## 4. USE-RATE ACCELERATION

Increasing the use rate can be an effective method of acceleration for some products. Use-rate acceleration may be appropriate for products such as electrical motors, relays and switches, paper copiers, printers, and home appliances such as toasters and washing machines. Also it is common practice to increase the cycling rate (or frequency) in fatigue testing. The manner in which the use rate is increased may depend on the product.

## 4.1 Simple Use-Rate Acceleration Models

There is a basic assumption underlying simple use-rate acceleration models. Useful life must be adequately modeled with cycles of operation as the time scale and cycling rate (or frequency) should not affect the cycles-to-failure distribution. This is reasonable if cycling simulates actual use and if the cycling frequency is low enough that test units return to steady state after each cycle (e.g., cool down).

In such simple situations, where the cycles-to-failure distribution does not depend on the cycling rate, we say that *reciprocity holds*. This implies that the underlying model for lifetime versus use rate is SAFT where $\mathcal{AF}(\text{UseRate}) = \text{UseRate}/\text{UseRate}_U$ is the factor by which the test is accelerated. For example, Nelson (1990, page 16) states that failure of rolling bearings can be accelerated by running them at three or more times the normal speed. Johnston et al. (1979) demonstrated that the cycles-to-failure of electrical insulation was shortened, approximately, by a factor of $\mathcal{AF}(412) = 412/60 \approx 6.87$ when the applied AC voltage in endurance tests was increased from 60 Hz to 412 Hz.

ALTs with increased use rate attempt to simulate actual use. Thus other environmental factors should be controlled to mimic actual use environments. If the cycling rate is too high, it can cause *reciprocity breakdown*. For example, it is necessary to have test units (such as a toaster) "cool down" between cycles of operation. Otherwise, heat buildup can cause the cycles-to-failure distribution to depend on the cycling rate.

## 4.2 Cycles to Failure Depends on Use Rate

Testing at higher frequencies could shorten test times but could also affect the cycles-to-failure distribution due to specimen heating or other effects. In some complicated situations, wear rate or degradation rate depends on cycling frequency. Also, a product may deteriorate in stand-by as well as during actual use. Reciprocity breakdown is known to occur, for example, for certain components in copying machines where components tend to last longer (in terms of cycles) when printing is done at higher rates. Dowling (1993, page 706) describes how increased cycle rate may affect the crack growth rate in per cycle fatigue testing. In such cases, the empirical power-rule relationship $\mathcal{AF}(\text{UseRate}) = (\text{UseRate}/\text{UseRate}_U)^p$ is often used, where $p$ can be estimated by testing at two or more use rates.

EXAMPLE 2 (*Increased cycling rate for low-cycle fatigue tests*). Fatigue life is typically measured in cycles to failure. To estimate low-cycle fatigue life of metal specimens, testing is done using cycling rates typically ranging between 10 Hz and 50 Hz (where 1 Hz is one stress cycle per second), depending on material type and available test equipment. At 50 Hz, accumulation of $10^6$ cycles would require about five hours of testing. Accumulation of $10^7$ cycles would require about two days and accumulation of $10^8$ cycles would require about 20 days. Higher frequencies are used in the study of high-cycle fatigue.

Some fatigue tests are conducted to estimate crack growth rates, often as a function of explanatory variables like stress and temperature. Such tests generally use rectangular compact tension test specimens containing a long slot cut normal to the centerline with a chevron machined into the end of the notch. Because the location of the chevron is a point of highest stress, a crack will initiate and grow from there. Other fatigue tests measure cycles to failure. Such tests use cylindrical dog-bone-shaped specimens. Again, cracks tend to initiate in the narrow part of the dog bone, although sometimes a notch is cut into the specimen to initiate the crack.

Cycling rates in fatigue tests are generally increased to a point where the desired response can still be measured without distortion. For both kinds of fatigue tests, the results are used as inputs to engineering models that predict the life of actual system components. The details of such models that are actually used in practice are usually proprietary, but are typified, for example, by Miner's rule (e.g., page 494 of Nelson, 1990) which uses results of tests in which specimens are tested at constant stress to predict life in which system components are exposed to varying stresses. Example 15.3 in Meeker and Escobar (1998) describes, generally, how results of fatigue tests on specimens are used to predict the reliability of a jet engine turbine disk.

There is a danger, however, that increased temperature due to increased cycling rate will affect the cycles-to-failure distribution. This is especially true if there are effects like creep-fatigue interaction (see Dowling, 1993, page 706, for further discussion). In another example, there was concern that changes in cycling rate would affect the distribution of lubricant on a rolling bearing surface. In particular, if $T$ is life in cycles and $T$ has a log-location-scale distribution with parameters $(\mu, \sigma)$, then $\mu = \beta_0 + \beta_1 \log(\text{cycles/unit time})$ where $\beta_0$ and $\beta_1$ can be estimated from data at two or more values of cycles/unit time.

## 5. USING TEMPERATURE TO ACCELERATE FAILURE MECHANISMS

It is sometimes said that high temperature is the enemy of reliability. Increasing temperature is one of the most commonly used methods to accelerate a failure mechanism.

### 5.1 Arrhenius Relationship for Reaction Rates

The Arrhenius relationship is a widely used model to describe the effect that temperature has on the rate of a simple chemical reaction. This relationship can be written as

$$(3) \qquad \mathcal{R}(\texttt{temp}) = \gamma_0 \exp\left(\frac{-E_a}{k \times \texttt{temp K}}\right)$$

where $\mathcal{R}$ is the reaction rate, and $\texttt{temp K} = \texttt{temp °C} + 273.15$ is thermodynamic temperature in kelvin (K), $k$ is either Boltzmann's constant or the universal gas constant and $E_a$ is the activation energy. The parameters $E_a$ and $\gamma_0$ are product or material characteristics. In applications involving electronic component reliability, Boltzmann's constant $k = 8.6171 \times 10^{-5} = 1/11605$ in units of electronvolt per kelvin (eV/K) is commonly used and in this case, $E_a$ has units of electronvolt (eV).

In the case of a simple one-step chemical reaction, $E_a$ would represent an activation energy that quantifies the minimum amount of energy needed to allow a certain chemical reaction to occur. In most applications involving temperature acceleration of a failure mechanism, the situation is much more complicated. For example, a chemical degradation process may have multiple steps operating in series or parallel, with each step having its own rate constant and activation energy. Generally, the hope is that the behavior of the more complicated process can be approximated, over the entire range of temperature of interest, by the Arrhenius relationship. This hope can be realized, for example, if there is a single step in the degradation process that is rate-limiting and thus, for all practical purposes, controls the rate of the entire reaction. Of course this is a strong assumption that in most practical applications is impossible to verify completely. In most accelerated test applications, it would be more appropriate to refer to $E_a$ in (3) as a *quasi-activation energy*.

### 5.2 Arrhenius Relationship Time-Acceleration Factor

The Arrhenius acceleration factor is

$$
\begin{aligned}
(4) \quad &\mathcal{AF}(\texttt{temp}, \texttt{temp}_U, E_a) \\
&= \frac{\mathcal{R}(\texttt{temp})}{\mathcal{R}(\texttt{temp}_U)} \\
&= \exp\left[E_a\left(\frac{11605}{\texttt{temp}_U\,\text{K}} - \frac{11605}{\texttt{temp K}}\right)\right].
\end{aligned}
$$

When $\texttt{temp} > \texttt{temp}_U$, $\mathcal{AF}(\texttt{temp}, \texttt{temp}_U, E_a) > 1$. When $\texttt{temp}_U$ and $E_a$ are understood to be, respectively, product use temperature and reaction-specific quasi activation energy, $\mathcal{AF}(\texttt{temp}) = \mathcal{AF}(\texttt{temp}, \texttt{temp}_U, E_a)$ will be used to denote a time-acceleration factor. The following example illustrates how one can assess approximate acceleration factors for a proposed accelerated test.

EXAMPLE 3 (*Adhesive-bonded power element*). Meeker and Hahn (1985) describe an adhesive-bonded power element that was designed for use at $\texttt{temp} = 50$ °C. Suppose that a life test of this element is to be conducted at $\texttt{temp} = 120$ °C. Also suppose that experience with this product suggested that $E_a$ can vary in the range $E_a = 0.4$ eV to $E_a = 0.6$ eV. Figure 3 gives the acceleration factors for the chemical reaction when testing the power element at 120 °C and quasi-activation energies of $E_a = (0.4, 0.5, 0.6)$ eV. The corresponding approximate acceleration factors at 120 °C are $\mathcal{AF}(120) = 12.9, 24.5$, and 46.4, respectively.

The Arrhenius relationship does not apply to all temperature acceleration problems and will be adequate over only a limited temperature range (depending on the particular application). Yet it is satisfactorily and widely used in many applications. Nelson (1990, page 76) comments that "...in certain applications (e.g., motor insulation), if the Arrhenius relationship ...does not fit the data, the data are suspect rather than the relationship."

### 5.3 Eyring Relationship Time-Acceleration Factor

The Arrhenius relationship (3) was discovered by Svante Arrhenius through empirical observation in the late 1800s. Eyring (e.g., Gladstone, Laidler and Eyring, 1941, or Eyring, 1980) gives physical theory describing the effect that temperature has on a reaction rate. Written in terms of a reaction rate, the

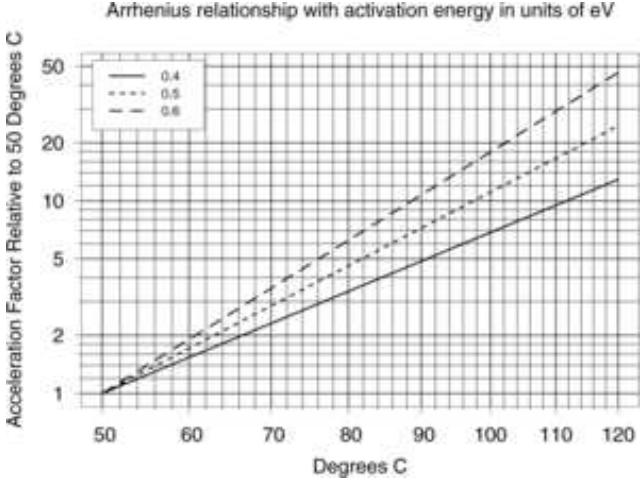

Fig. 3. *Time-acceleration factors as a function of temperature for the adhesive-bonded example with $E_a = (0.4, 0.5, 0.6)$ eV.*

Eyring relationship is

$$\mathcal{R}(\texttt{temp}) = \gamma_0 \times A(\texttt{temp}) \times \exp\left(\frac{-E_a}{k \times \texttt{temp K}}\right)$$

where $A(\texttt{temp})$ is a function of temperature depending on the specifics of the reaction dynamics and $\gamma_0$ and $E_a$ are constants (Weston and Schwarz, 1972, e.g., provides more detail). Applications in the literature have typically used $A(\texttt{temp}) = (\texttt{temp K})^m$ with a fixed value of $m$ ranging between $m = 0$ (Boccaletti et al., 1989, page 379), $m = 0.5$ (Klinger, 1991), to $m = 1$ (Nelson, 1990, page 100, and Mann, Schafer and Singpurwalla, 1974, page 436).

The Eyring relationship temperature acceleration factor is

$$\mathcal{AF}_{\text{Ey}}(\texttt{temp}, \texttt{temp}_U, E_a)$$
$$= \left(\frac{\texttt{temp K}}{\texttt{temp}_U \, K}\right)^m$$
$$\times \mathcal{AF}_{\text{Ar}}(\texttt{temp}, \texttt{temp}_U, E_a)$$

where $\mathcal{AF}_{\text{Ar}}(\texttt{temp}, \texttt{temp}_U, E_a)$ is the Arrhenius acceleration factor from (4). For use over practical ranges of temperature acceleration, and for practical values of $m$ not far from 0, the factor outside the exponential has relatively little effect on the acceleration factor and the additional term is often dropped in favor of the simpler Arrhenius relationship.

EXAMPLE 4 (*Eyring acceleration factor for a metallization failure mode*). An accelerated life test will be used to study a metallization failure mechanism for a solid-state electronic device. Experience with this type of failure mechanism suggests that the quasi-activation energy should be in the neighborhood of $E_a = 1.2$ eV. The usual operating junction temperature for the device is 90 °C. The Eyring acceleration factor for testing at 160 °C, using $m = 1$, is

$$\mathcal{AF}_{\text{Ey}}(160, 90, 1.2)$$
$$= \left(\frac{160 + 273.15}{90 + 273.15}\right) \times \mathcal{AF}_{\text{Ar}}(160, 90, 1.2)$$
$$= 1.1935 \times 491 = 586$$

where $\mathcal{AF}_{\text{Ar}}(160, 90, 1.2) = 491$ is the Arrhenius acceleration factor. We see that, for a *fixed* value of $E_a$, the Eyring relationship predicts, in this case, an acceleration that is 19% greater than the Arrhenius relationship. As explained below, however, this figure exaggerates the practical difference between these models.

When fitting models to limited data, the estimate of $E_a$ depends strongly on the assumed value for $m$ (e.g., 0 or 1). This dependency will compensate for and reduce the effect of changing the assumed value of $m$. Only with extremely large amounts of data would it be possible to adequately separate the effects of $m$ and $E_a$ using data alone. If $m$ can be determined accurately on the basis of physical considerations, the Eyring relationship could lead to better low-stress extrapolations. Numerical evidence shows that the acceleration factor obtained from the Eyring model assuming $m$ known, and estimating $E_a$ from the data, is monotone decreasing as a function of $m$. Then the Eyring model gives smaller acceleration factors and smaller extrapolation to use levels of temperature when $m > 0$. When $m < 0$, Arrhenius gives a smaller acceleration factor and a conservative extrapolation to use levels of temperature.

### 5.4 Reaction-Rate Acceleration for a Nonlinear Degradation Path Model

Some simple chemical degradation processes (first-order kinetics) might be described by the following path model:

$$\begin{aligned}(5) \quad & \mathcal{D}(t; \texttt{temp}) \\ & = \mathcal{D}_\infty \times \{1 - \exp[-\mathcal{R}_U \times \mathcal{AF}(\texttt{temp}) \times t]\}\end{aligned}$$

where $\mathcal{R}_U$ is the reaction rate at use temperature $\texttt{temp}_U$, $\mathcal{R}_U \times \mathcal{AF}(\texttt{temp})$ is the rate reaction at a general temperature $\texttt{temp}$, and for $\texttt{temp} > \texttt{temp}_U$, $\mathcal{AF}(\texttt{temp}) > 1$. Figure 4 shows this function for fixed $\mathcal{R}_U$, $E_a$ and $\mathcal{D}_\infty$, but at different temperatures. Note from (5) that when $\mathcal{D}_\infty > 0$, $\mathcal{D}(t)$ is increasing and failure occurs when $\mathcal{D}(t) > \mathcal{D}_\mathrm{f}$. For the example in Figure 4, however, $\mathcal{D}_\infty < 0$, $\mathcal{D}(t)$ is decreasing, and failure occurs when $\mathcal{D}(t) < \mathcal{D}_\mathrm{f} = -0.5$. In either case, equating $\mathcal{D}(T;\texttt{temp})$ to $\mathcal{D}_\mathrm{f}$ and solving for failure time gives

$$(6) \qquad T(\texttt{temp}) = \frac{T(\texttt{temp}_U)}{\mathcal{AF}(\texttt{temp})}$$

where

$$T(\texttt{temp}_U) = -\left(\frac{1}{\mathcal{R}_U}\right)\log\left(1 - \frac{\mathcal{D}_\mathrm{f}}{\mathcal{D}_\infty}\right)$$

is failure time at use conditions. Faster degradation shortens time to any particular definition of failure (e.g., crossing $\mathcal{D}_\mathrm{f}$ or some other specified level) by a *scale factor* that depends on temperature. Thus changing temperature is similar to changing the units of time. Consequently, the time-to-failure distributions at $\texttt{temp}_U$ and $\texttt{temp}$ are related by

$$(7) \qquad \begin{aligned} &\Pr[T(\texttt{temp}_U) \le t] \\ &\quad = \Pr[T(\texttt{temp}) \le t/\mathcal{AF}(\texttt{temp})]. \end{aligned}$$

Equations (6) and (7) are forms of the scale-accelerated failure-time (SAFT) model introduced in Section 3.2.

With a SAFT model, for example, if $T(\texttt{temp}_U)$ (time at use or some other baseline temperature) has a log-location-scale distribution with parameters $\mu_U$ and $\sigma$, then

$$\Pr[T \le t; \texttt{temp}_U] = \Phi\left[\frac{\log(t) - \mu_U}{\sigma}\right].$$

At any other temperature,

$$\Pr[T \le t; \texttt{temp}] = \Phi\left[\frac{\log(t) - \mu}{\sigma}\right]$$

where

$$\mu = \mu(x) = \mu_U - \log[\mathcal{AF}(\texttt{temp})] = \beta_0 + \beta_1 x,$$

$x = 11605/(\texttt{temp K})$, $x_U = 11605/(\texttt{temp}_U \text{K})$, $\beta_1 = E_a$ and $\beta_0 = \mu_U - \beta_1 x_U$. LuValle, Welsher and Svoboda (1988) and Klinger (1992) describe more general physical/chemical degradation model characteristics needed to assure that the SAFT property holds.

EXAMPLE 5 (*Time acceleration for Device-A*). Hooper and Amster (1990) analyze the temperature-accelerated life test data on a particular kind of electronic device that is identified here as Device-A. The data are given in Meeker and Escobar (1998, page 637). The purpose of the experiment was to determine if Device-A would meet its failure-rate objective through 10,000 hours and 30,000 hours at its nominal operating ambient temperature of 10 °C. Figure 5 shows the censored life data and the Arrhenius-lognormal ML fit of the distribution quantiles versus temperature, describing the relationship between life and temperature. There were 0 failure out of 30 units tested at 10 °C, 10 out of 100 at 40 °C, 9 out of 20 at 60 °C, and 14 out of 15 at 80 °C. The censored observations are denoted in Figure 5 by $\Delta$. The life-temperature relationship plots as a family of straight lines because temperature is plotted on an Arrhenius axis and life is plotted on a log axis. The densities are normal densities because the lognormal life distributions are plotted on a log axis.

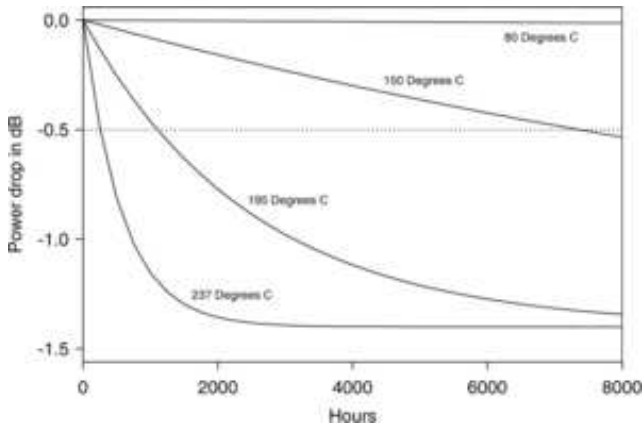

FIG. 4. *Nonlinear degradation paths at different temperatures with a SAFT relationship.*

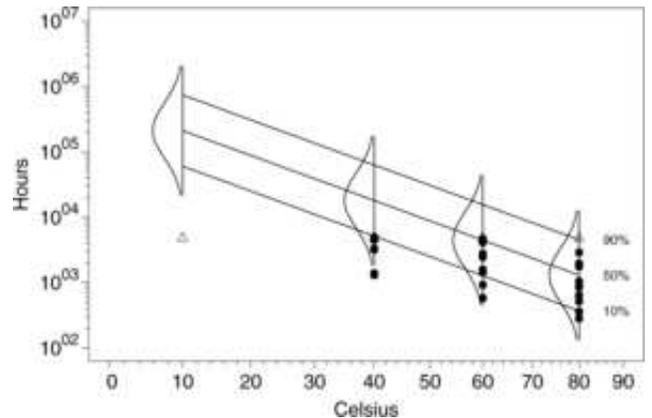

FIG. 5. *Arrhenius-lognormal model fitted to the Device-A data. Censored observations are indicated by $\Delta$.*

## 5.5 Examples where the Arrhenius Model is not Appropriate

As described in Section 5.1, strictly speaking, the Arrhenius relationship will describe the rate of a chemical reaction only under special circumstances. It is easy to construct examples where the Arrhenius model does not hold. For example, if there is more than one competing chemical reaction and those chemical reactions have different activation energies, the Arrhenius model will not describe the rate of the overall chemical reaction.

EXAMPLE 6 (*Acceleration of parallel chemical reactions*). Consider the chemical degradation path model having two separate reactions contributing to failure and described by

$$
\begin{aligned}
\mathcal{D}(t; &\texttt{temp}) \\
&= \mathcal{D}_{1\infty} \times \{1 - \exp[-\mathcal{R}_{1U} \times \mathcal{AF}_1(\texttt{temp}) \times t]\} \\
&\quad + \mathcal{D}_{2\infty} \\
&\quad \times \{1 - \exp[-\mathcal{R}_{2U} \times \mathcal{AF}_2(\texttt{temp}) \times t]\}.
\end{aligned}
$$

Here $\mathcal{R}_{1U}$ and $\mathcal{R}_{2U}$ are the use-condition rates of the two parallel reactions contributing to failure. Suppose that the Arrhenius relationship can be used to describe temperature dependence for these rates, providing acceleration functions $\mathcal{AF}_1(\texttt{temp})$ and $\mathcal{AF}_2(\texttt{temp})$. Then, unless $\mathcal{AF}_1(\texttt{temp}) = \mathcal{AF}_2(\texttt{temp})$ for all $\texttt{temp}$, this degradation model does *not* lead to a SAFT model. Intuitively, this is because temperature affects the two degradation processes differently, inducing a nonlinearity into the acceleration function relating times at two different temperatures.

To obtain useful extrapolation models for degradation processes having more than one step, each with its own rate constant, it is, in general, necessary to have adequate models for the important individual steps. For example, when the individual processes can be observed, it may be possible to estimate the effect that temperature (or other accelerating variables) has on each of the rate constants.

## 5.6 Other Units for Activation Energy

The discussion and examples of the Arrhenius and Eyring relationships in Sections 5.2–5.4 used units of electronvolt for $E_a$ and electronvolt per kelvin for $k$. These units for the Arrhenius model are used most commonly in applications involving electronics. In other areas of application (e.g., degradation of organic materials such as paints and coatings, plastics, food and pharmaceuticals), it is more common to see Boltzmann's constant $k$ in units of electronvolt replaced with the universal gas constant in other units. For example, the gas constant is commonly given in units of kilojoule per mole kelvin [i.e., R = 8.31447 kJ/(mol·K)]. In this case, $E_a$ is activation energy in units of kilojoule per mole (kJ/mol). The corresponding Arrhenius acceleration factor is

$$
\begin{aligned}
\mathcal{AF}(&\texttt{temp}, \texttt{temp}_U, E_a) \\
&= \exp\left[E_a\left(\frac{120.27}{\texttt{temp}_U\,\text{K}} - \frac{120.27}{\texttt{temp}\,\text{K}}\right)\right].
\end{aligned}
$$

The universal gas constant can also be expressed in units of kilocalorie per mole kelvin, kcal/(mol·K) [i.e., R = 1.98588 kcal/(mol·K)]. In this case, $E_a$ is in units of kilocalorie per mole (kcal/mol). The corresponding Arrhenius acceleration factor is

$$
\begin{aligned}
\mathcal{AF}(&\texttt{temp}, \texttt{temp}_U, E_a) \\
&= \exp\left[E_a\left(\frac{503.56}{\texttt{temp}_U\,\text{K}} - \frac{503.56}{\texttt{temp}\,\text{K}}\right)\right].
\end{aligned}
$$

It is also possible to use units of kJ/(mol·K) and kcal/(mol·K) for the $E_a$ coefficient in the Eyring model.

Although $k$ is standard notation for Boltzmann's constant and R is standard notation for the universal gas constant, we use $k$ to denote either of these in the Arrhenius relationship.

## 5.7 Temperature Cycling

Some failure modes are caused by temperature cycling. In particular, temperature cycling causes thermal expansion and contraction which can induce mechanical stresses. Some failure modes caused by thermal cycling include:

- Power on/off cycling of electronic equipment can damage integrated circuit encapsulement and solder joints.
- Heat generated by take-off power-thrust in jet engines can cause crack initiation and growth in fan disks.
- Power-up/power-down cycles can cause cracks in nuclear power plant heat exchanger tubes and turbine generator components.
- Temperature cycling can lead to delamination in inkjet printhead components.

As in fatigue testing, it is possible to accelerate thermal cycling failure modes by increasing either the frequency or amplitude of the cycles (increasing amplitude generally increases mechanical stress). The most commonly used model for acceleration of thermal cycling is the Coffin–Manson relationship which says that the number of cycles to failure is

$$N = \frac{\delta}{(\Delta\texttt{temp})^{\beta_1}}$$

where $\Delta\texttt{temp}$ is the temperature range and $\delta$ and $\beta_1$ are properties of the material and test setup. This power-rule relationship explains the effect that temperature range has on the thermal-fatigue life cycles-to-failure distribution. Nelson (1990, page 86) suggests that for some metals, $\beta_1 \approx 2$ and that for plastic encapsulements used for integrated circuits, $\beta_1 \approx 5$. The Coffin–Manson relationship was originally developed as an empirical model to describe the effect of temperature cycling on the failure of components in the hot part of a jet engine. See Nelson (1990, page 86) for further discussion and references.

Letting $T$ be the random number of cycles to failure (e.g., $T = N\varepsilon$ where $\varepsilon$ is a random variable). The acceleration factor at $\Delta\texttt{temp}$, relative to $\Delta\texttt{temp}_U$, is

$$\mathcal{AF}(\Delta\texttt{temp}) = \frac{T(\Delta\texttt{temp}_U)}{T(\Delta\texttt{temp})} = \left(\frac{\Delta\texttt{temp}}{\Delta\texttt{temp}_U}\right)^{\beta_1}.$$

There may be a $\Delta\texttt{temp}$ threshold below which little or no fatigue damage is done during thermal cycling.

Empirical evidence has shown that the effect of temperature cycling can depend importantly on $\texttt{temp}_{\max}$ K, the maximum temperature in the cycling (e.g., if $\texttt{temp}_{\max}$ K is more than 0.2 or 0.3 times a metal's melting point). The cycles-to-failure distribution for temperature cycling can also depend on the cycling rate (e.g., due to heat buildup). An empirical extension of the Coffin–Manson relationship that describes such dependencies is

$$N = \frac{\delta}{(\Delta\texttt{temp})^{\beta_1}} \times \frac{1}{(\texttt{freq})^{\beta_2}} \times \exp\left(\frac{E_a \times 11605}{\texttt{temp}_{\max}\,\text{K}}\right),$$

where $\texttt{freq}$ is the cycling frequency and $E_a$ is a quasi-activation energy.

As with all acceleration models, caution must be used when using such a model outside the range of available data and past experience.

## 6. USING HUMIDITY TO ACCELERATE REACTION RATES

Humidity is another commonly used accelerating variable, particularly for failure mechanisms involving corrosion and certain kinds of chemical degradation.

EXAMPLE 7 (*Accelerated life test of a printed wiring board*). Figure 6 shows data from an ALT of printed circuit boards, illustrating the use of humidity as an accelerating variable. This is a subset of the larger experiment described by LuValle, Welsher and Mitchell (1986), involving acceleration with temperature, humidity and voltage. A table of the data is given in Meeker and LuValle (1995) and in Meeker and Escobar (1998). Figure 6 shows clearly that failures occur earlier at higher levels of humidity.

Vapor density measures the amount of water vapor in a volume of air in units of mass per unit volume. Partial vapor pressure (sometimes simply referred to as "vapor pressure") is closely related and measures that part of the total air pressure exerted by the water molecules in the air. Partial vapor pressure is approximately proportional to vapor density. The partial vapor pressure at which molecules are evaporating and condensing from the surface of water at the same rate is the saturation vapor pressure. For a fixed amount of moisture in the air, saturation vapor pressure increases with temperature.

Relative humidity is usually defined as

$$\texttt{RH} = \frac{\text{Vapor Pressure}}{\text{Saturation Vapor Pressure}}$$

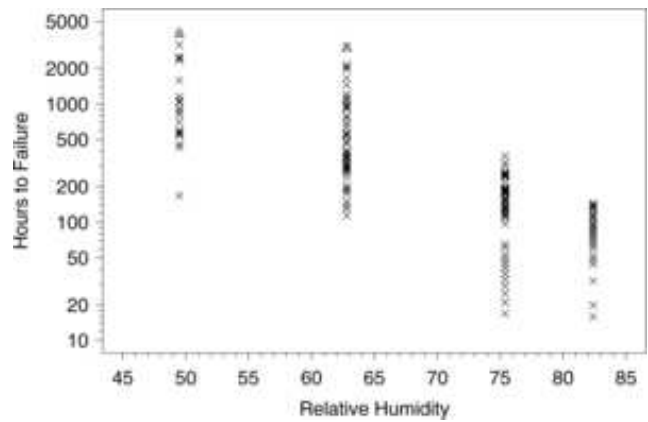

FIG. 6. *Scatterplot of printed circuit board accelerated life test data. Censored observations are indicated by $\Delta$. There are 48 censored observations at 4078 hours in the 49.5% RH test and 11 censored observations at 3067 hours in the 62.8% RH test.*

and is commonly expressed as a percent. For most failure mechanisms, physical/chemical theory suggests that RH is the appropriate scale in which to relate reaction rate to humidity especially if temperature is also to be used as an accelerating variable (Klinger, 1991).

A variety of different humidity models (mostly empirical but a few with some physical basis) have been suggested for different kinds of failure mechanisms. Much of this work has been motivated by concerns about the effect of environmental humidity on plastic-packaged electronic devices. Humidity is also an important factor in the service-life distribution of paints and coatings. In most test applications where humidity is used as an accelerating variable, it is used in conjunction with temperature. For example, Peck (1986) presents data and models relating life of semiconductor electronic components to humidity and temperature. See also Peck and Zierdt (1974) and Joyce et al. (1985). Gillen and Mead (1980) describe a kinetic approach for modeling accelerated aging data. LuValle, Welsher and Mitchell (1986) describe the analysis of time-to-failure data on printed circuit boards that have been tested at higher than usual temperature, humidity and voltage. They suggest ALT models based on the physics of failure. Chapter 2 of Nelson (1990) and Boccaletti et al. (1989) review and compare a number of different humidity models.

The Eyring/Arrhenius temperature-humidity acceleration relationship in the form of (14) uses $x_1 = 11605/\texttt{temp}\,\text{K}$, $x_2 = \log(\texttt{RH})$ and $x_3 = x_1 x_2$ where $\texttt{RH}$ is relative humidity, expressed as a proportion. An alternative humidity relationship suggested by Klinger (1991), on the basis of a simple kinetic model for corrosion, uses the term $x_2 = \log[\texttt{RH}/(1 - \texttt{RH})]$ (a logistic transformation) instead.

In most applications where it is used as an accelerating variable, higher humidity increases degradation rates and leads to earlier failures. In applications where drying is the failure mechanism, however, an artificial environment with lower humidity can be used to accelerate a test.

# 7. ACCELERATION MODEL FOR PHOTODEGRADATION

Many organic compounds degrade chemically when exposed to ultraviolet (UV) radiation. Such degradation is known as photodegradation. This section describes models that have been used to study photodegradation and that are useful when analyzing data from accelerated photodegradation tests. Many of the ideas in this section originated from early research into the effects of light on photographic emulsions (e.g., James, 1977) and the effect that UV exposure has on causing skin cancer (e.g., Blum, 1959). Important applications include prediction of service life of products exposed to UV radiation (outdoor weathering) and fiber-optic systems.

## 7.1 Time Scale and Model for Total Effective UV Dosage

As described in Martin et al. (1996), the appropriate time scale for photodegradation is the total (i.e., cumulative) effective UV dosage, denoted by $D_{\text{Tot}}$. Intuitively, this total effective dosage can be thought of as the cumulative number of photons absorbed into the degrading material and that cause chemical change. The total effective UV dosage at real time $t$ can be expressed as

$$(8) \qquad D_{\text{Tot}}(t) = \int_0^t D_{\text{Inst}}(\tau)\,d\tau$$

where the instantaneous effective UV dosage at real time $\tau$ is

$$D_{\text{Inst}}(\tau) = \int_{\lambda_1}^{\lambda_2} D_{\text{Inst}}(\tau, \lambda)\,d\lambda$$

$$(9) \qquad = \int_{\lambda_1}^{\lambda_2} E_0(\lambda, \tau)$$
$$\times \{1 - \exp[-A(\lambda)]\}\phi(\lambda)\,d\lambda.$$

Here $E_0(\lambda, \tau)$ is the spectral irradiance (or intensity) of the light source at time $\tau$ (both artificial and natural light sources have potentially time-dependent mixtures of light at different wavelengths, denoted by $\lambda$), $[1 - \exp(-A(\lambda))]$ is the spectral absorbance of the material being exposed (damage is caused only by photons that are absorbed into the material), and $\phi(\lambda)$ is a quasi-quantum efficiency of the absorbed radiation (allowing for the fact that photons at certain wavelengths have a higher probability of causing damage than others). The functions $E_0$ and $A$ in the integrand of (9) can either be measured directly or estimated from data and the function $\phi(\lambda)$ can be estimated from data. A simple log-linear model is commonly used to describe quasi-quantum efficiency as a function of wavelength. That is,

$$\phi(\lambda) = \exp(\beta_0 + \beta_1 \lambda).$$

The integrals over wavelength, like that in (9), are typically taken over the UV-B band (290 nm to 320 nm), as this is the range of wavelengths over which both $\phi(\lambda)$ and $E_0(\lambda, t)$ are importantly different from 0. Longer wavelengths (in the UV-A band) are not terribly harmful to organic materials [$\phi(\lambda) \approx 0$]. Shorter wavelengths (in the UV-C band) have more energy, but are generally filtered out by ozone in the atmosphere [$E_0(\lambda, t) \approx 0$].

### 7.2 Additivity

Implicit in the model (9) is the assumption of additivity. Additivity implies, in this setting, that the photo-effectiveness of a source is equal to the sum of the effectiveness of its spectral components. This part of the model makes it relatively easy to conduct exposure tests with specific combinations of wavelengths [e.g., by using selected band-pass filters to define $E_0(\lambda, \tau)$ functions as levels of spectral intensity in an experiment] to estimate the quasi-quantum efficiency as a function of $\lambda$. Then the total dosage model in (9) can be used to predict photodegradation under other combinations of wavelengths [i.e., for other $E_0(\lambda, \tau)$ functions].

### 7.3 Reciprocity and Reciprocity Breakdown

The intuitive idea behind reciprocity in photodegradation is that the time to reach a certain level of degradation is inversely proportional to the rate at which photons attack the material being degraded. Reciprocity breakdown occurs when the coefficient of proportionality changes with light intensity. Although reciprocity provides an adequate model for some degradation processes (particularly when the range of intensities used in experimentation and actual applications is not too broad), some examples have been reported in which there is reciprocity breakdown (e.g., Blum, 1959, and James, 1977).

Light intensity can be affected by filters. Sunlight is filtered by the earth's atmosphere. In laboratory experiments, different neutral density filters are used to reduce the amount of light passing to specimens (without having an important effect on the wavelength spectrum), providing an assessment of the degree of reciprocity breakdown. Reciprocity implies that the effective time of exposure is

$$d(t) = \text{CF} \times \text{D}_{\text{Tot}}(t)$$
$$= \text{CF} \times \left[ \int_0^t \int_{\lambda_1}^{\lambda_2} \text{D}_{\text{Inst}}(\tau, \lambda) \, d\lambda \, d\tau \right]$$

where CF is an "acceleration factor." For example, commercial outdoor test exposure sites use mirrors to concentrate light to achieve, say, "5 Suns" acceleration or CF = 5. A 50% neutral density filter in a laboratory experiment will provide deceleration corresponding to CF = 0.5.

When there is evidence of reciprocity breakdown, the effective time of exposure is often modeled, empirically, by

$$\begin{aligned} (10) \quad d(t) &= (\text{CF})^p \times \text{D}_{\text{Tot}}(t) \\ &= (\text{CF})^p \times \left[ \int_0^t \int_{\lambda_1}^{\lambda_2} \text{D}_{\text{Inst}}(\tau, \lambda) \, d\lambda \, d\tau \right]. \end{aligned}$$

Model (10) has been shown to fit data well and experimental work in the photographic literature suggests that when there is reciprocity breakdown, the value of $p$ does not depend strongly, if at all, on the wavelength $\lambda$. A statistical test for $p = 1$ can be used to assess the reciprocity assumption.

### 7.4 Model for Photodegradation and UV Intensity

Degradation (or damage) $D(t)$ at time $t$ depends on environmental variables like UV, temp and RH, that may vary over time, say according to a multivariable profile $\xi(t) = [\text{UV}, \text{temp}, \text{RH}, \ldots]$. Laboratory tests are conducted in well-controlled environments, usually holding these variables constant (although sometimes such variables are purposely changed during an experiment, as in step-stress accelerated tests). Interest often centers, however, on life in a variable environment. Figure 7 shows some typical sample paths (for FTIR peak at 1510 cm$^{-1}$, representing benzene ring mass loss) for several specimens of an epoxy exposed to UV radiation using a band-pass filter with a nominal center at 306 nm. Separate paths are shown for each combination of (10, 40, 60, 100)% neutral density filters and 45 °C and 55 °C, as a function of total (cumulative) absorbed UV-B dosage. These sample paths might be modeled by a given functional form,

$$D(t) = g(z), \quad z = \log[d(t)] - \mu,$$

where $z$ is scaled time and $g(z)$ would usually be suggested by knowledge of the kinetic model (e.g., linear for zeroth-order kinetics and exponential for first-order kinetics), although empirical curve fitting may be adequate for purposes where the amount of extrapolation in the time dimension is not large. As in SAFT models, $\mu$ can be modeled as a function of explanatory variables like temperature and humidity when these variables affect the degradation rate.

## 8. VOLTAGE AND VOLTAGE-STRESS ACCELERATION

Increasing voltage or voltage stress (electric field) is another commonly used method to accelerate failure of electrical materials and components like light bulbs, capacitors, transformers, heaters and insulation.

Voltage quantifies the amount of force needed to move an electric charge between two points. Physically, voltage can be thought of as the amount of pressure behind an electrical current. Voltage stress quantifies voltage per unit of thickness across a dielectric and is measured in units of volt/thickness (e.g., V/mm or kV/mm).

### 8.1 Voltage Acceleration Mechanisms

Depending on the failure mode, higher voltage stress can:

- accelerate failure-causing electrochemical reactions or the growth of failure-causing discontinuities in the dielectric material.
- increase the voltage stress relative to dielectric strength of a specimen. Units at higher stress will tend to fail sooner than those at lower stress.

Sometimes one or the other of these effects will be the primary cause of failure. In other cases, both effects will be important.

EXAMPLE 8 (*Accelerated life test of insulation for generator armature bars*). Doganaksoy, Hahn and Meeker (2003) discuss an ALT for a new mica-based insulation design for generator armature bars (GABs). Degradation of an organic binder in the insulation causes a decrease in voltage strength and this was the primary cause of failure in the insulation. The insulation was designed for use at a voltage stress of 120 V/mm. Voltage-endurance tests were conducted on 15 electrodes at each of five accelerated voltage levels between 170 V/mm and 220 V/mm (i.e., a total of 75 electrodes). Each test was run for 6480 hours at which point 39 of the electrodes had not yet failed. Table 1 gives the data from these tests. The insulation engineers were interested in the 0.01 and 0.05 quantiles of lifetime at the use condition of 120 V/mm. Figure 8 plots the insulation lifetimes against voltage stress.

### 8.2 Inverse Power Relationship

The inverse power relationship is frequently used to describe the effect that stresses like voltage and pressure have on lifetime. Voltage is used in the following discussion. When the thickness of a dielectric material or insulation is constant, voltage is proportional to voltage stress. Let volt denote voltage and let $\mathtt{volt}_U$ be the voltage at use conditions. The lifetime at stress level volt is given by

$$T(\mathtt{volt}) = \frac{T(\mathtt{volt}_U)}{\mathcal{AF}(\mathtt{volt})} = \left(\frac{\mathtt{volt}}{\mathtt{volt}_U}\right)^{\beta_1} T(\mathtt{volt}_U)$$

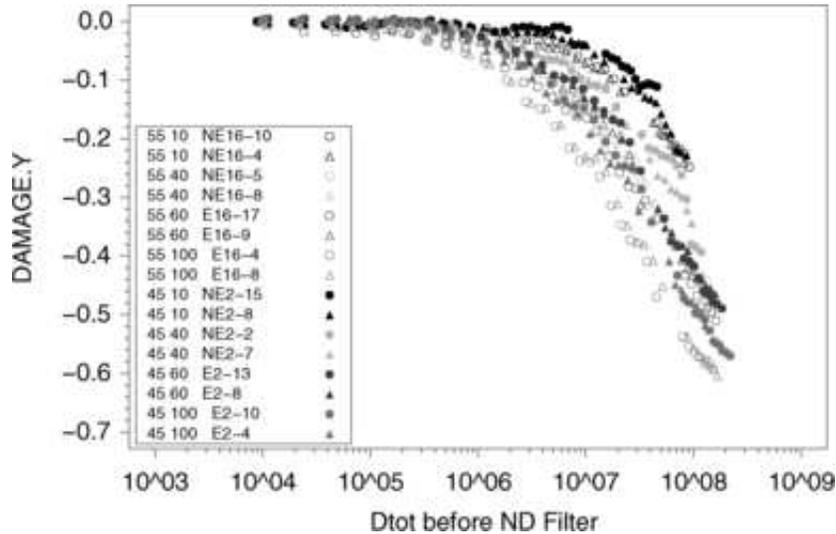

FIG. 7. *Sample paths for wave number 1510 $cm^{-1}$ and band-pass filter with nominal center at 306 nm for different combinations of temperature (45 and 55 °C) and neutral density filter [passing (10, 40, 60 and 100)% of photons across the UV-B spectrum].*

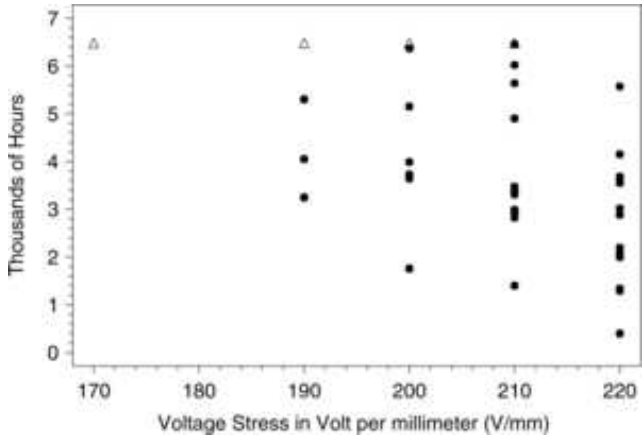

FIG. 8. *GAB insulation data. Scatterplot of life versus voltage. Censored observations are indicated by △.*

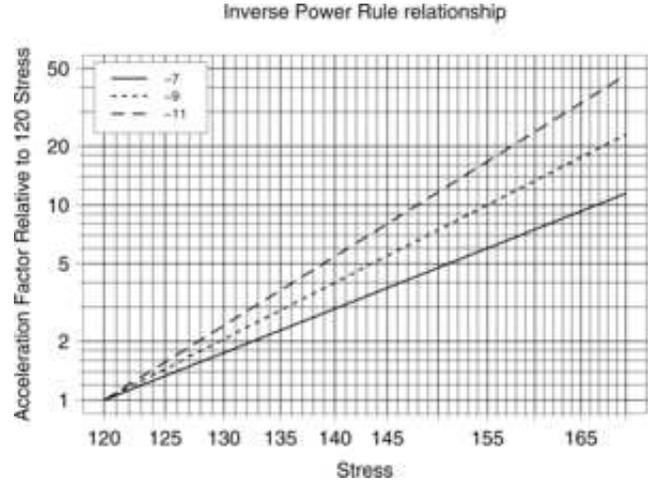

FIG. 9. *Time-acceleration factor as a function of voltage stress and exponent* $-\beta_1 = -7, -9, -11.$

where $\beta_1$, in general, is negative. The model has SAFT form with acceleration factor

$$
\begin{aligned}
\mathcal{AF}(\texttt{volt}) &= \mathcal{AF}(\texttt{volt}, \texttt{volt}_U, \beta_1) \\
&= \frac{T(\texttt{volt}_U)}{T(\texttt{volt})} \\
&= \left(\frac{\texttt{volt}}{\texttt{volt}_U}\right)^{-\beta_1}.
\end{aligned}
\tag{11}
$$

If $T(\texttt{volt}_U)$ has a log-location-scale distribution with parameters $\mu_U$ and $\sigma$, then $T(\texttt{volt})$ also has a log-location-scale distribution with $\mu = \beta_0 + \beta_1 x$, where $x_U = \log(\texttt{volt}_U)$, $x = \log(\texttt{volt})$, $\beta_0 = \mu_U - \beta_1 x_U$ and $\sigma$ does not depend on $x$.

EXAMPLE 9 (*Time acceleration for GAB insulation*). For the GAB insulation data in Example 8, an estimate for $\beta_1$ is $\widehat{\beta}_1 = -9$ (methods for computing such estimates are described in Meeker and Escobar, 1998, Chapter 19). Recall that the design voltage stress is $\texttt{volt}_U = 120$ V/mm and consider testing at $\texttt{volt} = 170$ V/mm. Thus, using $\beta_1 = \widehat{\beta}_1$, $\mathcal{AF}(170) = (170/120)^9 \approx 23$. Thus by increasing voltage stress from 120 V/mm to 170 V/mm, one estimates that lifetime is shortened by a factor of $1/\mathcal{AF}(170) \approx 1/23 = 0.04$. Figure 9 plots $\mathcal{AF}$ versus $\texttt{volt}$ for $\beta_1 = -7, -9, -11$. Using direct computations or from the plot, one obtains $\mathcal{AF}(170) \approx 11$ for $\beta_1 = -7$ and $\mathcal{AF}(170) \approx 46$ for $\beta_1 = -11$.

EXAMPLE 10 (*Accelerated life test of a mylar-polyurethane insulation*). Meeker and Escobar (1998, Section 19.3) reanalyzed ALT data from a special type of mylar-polyurethane insulation used in high-performance electromagnets. The data, originally from Kalkanis and Rosso (1989), give time to dielectric breakdown of units tested at (100.3, 122.4, 157.1, 219.0, 361.4) kV/mm. The purpose of the ALT was to evaluate the reliability of the insulating structure and to estimate the life distribution at system design voltages, assumed to be 50 kV/mm. Figure 10 shows that failures occur much sooner at high voltage stress. Except for the 361.4 kV/mm data, the relationship between log life and log voltage ap-pears to be approximately linear. Meeker and Escobar (1998), in their reanalysis, omitted the 361.4 kV/mm data because it is clear that a new failure mode had manifested itself at this highest level of voltage stress. Insulation engineers have suggested to us that the new failure mode was likely caused by thermal buildup that was not important at lower levels of voltage stress.

TABLE 1
*GAB insulation data*

| Voltage stress (V/mm) | Lifetime (thousand hours) |
| --- | --- |
| 170 | 15 censored[a] |
| 190 | 3.248, 4.052, 5.304, 12 censored[a] |
| 200 | 1.759, 3.645, 3.706, 3.726, 3.990, 5.153, 6.368, 8 censored[a] |
| 210 | 1.401, 2.829, 2.941, 2.991, 3.311, 3.364, 3.474, 4.902, 5.639, 6.021, 6.456, 4 censored[a] |
| 220 | 0.401, 1.297, 1.342, 1.999, 2.075, 2.196, 2.885, 3.019, 3.550, 3.566, 3.610, 3.659, 3.687, 4.152, 5.572 |

[a]Units were censored at 6.480 thousand hours.

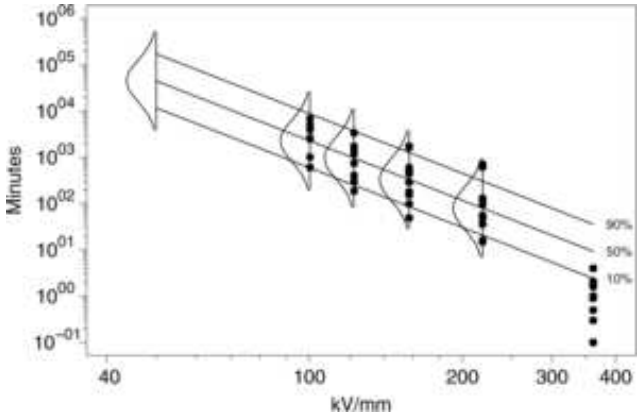

Fig. 10. *Inverse power relationship-lognormal model fitted to the mylar-polyurethane data (also showing the 361.4 kV/mm data omitted from the ML estimation).*

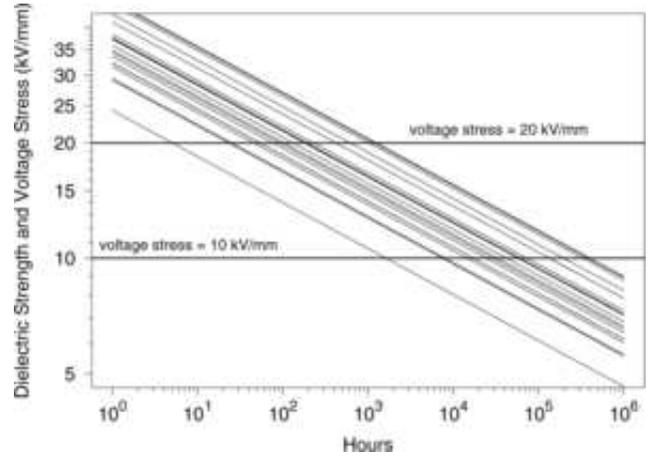

Fig. 11. *Dielectric strength degrading over time, relative to voltage-stress levels (horizontal lines).*

## 8.3 Physical Motivation for the Inverse Power Relationship for Voltage-Stress Acceleration

The inverse power relationship is widely used to model life as a function of pressure-like accelerating variables (e.g., stress, pressure, voltage stress). This relationship is generally considered to be an empirical model because it has no formal basis from knowledge of the physics/chemistry of the modeled failure modes. It is commonly used because engineers have found, over time, that it often provides a useful description of certain kinds of AT data.

This section presents a simple physical motivation for the inverse power relationship for voltage-stress acceleration under constant temperature situations. Section 9.2 describes a more general model for voltage acceleration involving a combination of temperature and voltage acceleration.

This discussion here is for insulation. The ideas extend, however, to other dielectric materials, products and devices like insulating fluids, transformers, capacitors, adhesives, conduits and containers that can be modeled by a stress-strength interference model.

In applications, an insulator should not conduct an electrical current. An insulator has a characteristic dielectric strength which can be expected to be random from unit to unit. The dielectric strength of an insulation specimen operating in a specific environment at a specific voltage may degrade with time. Figure 11 shows a family of simple curves to model degradation and unit-to-unit variability in dielectric strength over time. The unit-to-unit variability could be caused, for example, by materials or manufacturing variability. The horizontal lines represent voltage-stress levels that might be present in actual operation or in an accelerated test. When a specimen's dielectric strength falls below the applied voltage stress, there will be flash-over, a short circuit, or other failure-causing damage to the insulation. Analytically, suppose that degrading dielectric strength at age $t$ can be expressed as

$$\mathcal{D}(t) = \delta_0 \times t^{1/\beta_1}.$$

Here, as in Section 5.4, failure occurs when $\mathcal{D}(t)$ crosses $\mathcal{D}_f$, the applied voltage stress, denoted by volt. In Figure 11, the unit-to-unit variability is in the $\delta_0$ parameter. Equating $\mathcal{D}(T)$ to volt and solving for failure time $T$ gives

$$T(\texttt{volt}) = \left(\frac{\texttt{volt}}{\delta_0}\right)^{\beta_1}.$$

Then the acceleration factor for volt versus $\texttt{volt}_U$ is

$$\mathcal{AF}(\texttt{volt}) = \mathcal{AF}(\texttt{volt}, \texttt{volt}_U, \beta_1)$$
$$= \frac{T(\texttt{volt}_U)}{T(\texttt{volt})}$$
$$= \left(\frac{\texttt{volt}}{\texttt{volt}_U}\right)^{-\beta_1}$$

which is an inverse power relationship, as in (11).

To extend this model, suppose that higher voltage also leads to an increase in the degradation rate and that this increase is described with the degradation model

$$\mathcal{D}(t) = \delta_0 [\mathcal{R}(\texttt{volt}) \times t]^{1/\gamma_1}$$

where

$$\mathcal{R}(\texttt{volt}) = \gamma_0 \exp[\gamma_2 \log(\texttt{volt})].$$

Suppose failure occurs when $\mathcal{D}(t)$ crosses $\mathcal{D}_f$, the applied voltage stress, denoted by $\texttt{volt}$. Then equating $\mathcal{D}(T)$ to $\texttt{volt}$ and solving for failure time $T$ gives the failure time

$$T(\texttt{volt}) = \frac{1}{\mathcal{R}(\texttt{volt})} \left( \frac{\texttt{volt}}{\delta_0} \right)^{\gamma_1}.$$

Then the ratio of failure times at $\texttt{volt}_U$ versus $\texttt{volt}$ is the acceleration factor

$$\mathcal{AF}(\texttt{volt}) = \frac{T(\texttt{volt}_U)}{T(\texttt{volt})} = \left( \frac{\texttt{volt}}{\texttt{volt}_U} \right)^{\gamma_2 - \gamma_1},$$

which is again an inverse power relationship with $\beta_1 = \gamma_1 - \gamma_2$.

This motivation for the inverse power relationship described here is not based on any fundamental understanding of what happens to the insulating material at the molecular level over time. As we describe in Section 11, the use of such fundamental understanding could provide a better, more credible model for extrapolation.

### 8.4 Other Inverse Power Relationships

The inverse power relationship is also commonly used for other accelerating variables including pressure, cycling rate, electric current, stress and humidity. Some examples are given in Section 9.

### 8.5 A More General Empirical Power Relationship: Box–Cox Transformations

As shown in Section 8.2, the inverse power relationship induces a log-transformation in $\texttt{volt}$ giving the model $\mu = \beta_0 + \beta_1 x$, where $x = \log(\texttt{volt})$. There might be other transformations of $\texttt{volt}$ that could provide a better description of the data. A general, and useful, approach is to expand the formulation of the model by adding a parameter or parameters and investigating the effect of perturbing the added parameter(s), to see the effect on answers to questions of interest. Here this approach is used to expand the inverse power relationship model.

Suppose that $X_1$ is a positive accelerating variable and $\mathbf{X}_2$ is a collection of other explanatory variables, some of which might be accelerating variables. Consider the model $\mu = \beta_0 + \beta_1 X_1 + \boldsymbol{\beta}_2' \mathbf{X}_2$, where the $\beta$'s are unknown parameters. We start by replacing $X_1$ with the more general Box–Cox transformation (Box and Cox, 1964) on $X_1$. In particular, we fit the model

$$\mu = \gamma_0 + \gamma_1 W_1 + \boldsymbol{\gamma}_2' \mathbf{X}_2$$

where the $\gamma$'s are unknown parameters and

$$(12) \qquad W_1 = \begin{cases} \dfrac{X_1^\lambda - 1}{\lambda}, & \lambda \neq 0, \\ \log(X_1), & \lambda = 0. \end{cases}$$

The Box–Cox transformation (Box and Cox, 1964) was originally proposed as a simplifying transformation for a response variable. Transformation of accelerating and explanatory variables, however, provides a convenient extension of the accelerating modeling choices. The Box–Cox transformation includes all the power transformations and because $W_1$ is a continuous function of $\lambda$, (12) provides a continuum of transformations for possible evaluation and model assessment. The Box–Cox transformation parameter $\lambda$ can be varied over some range of values (e.g., $-1$ to $2$) to see the effect of different voltage-life relationships on the fitted model and inferences of interest. The results from the analysis can be displayed in a number of different ways.

For fixed $\mathbf{X}_2$, the Box–Cox transformation model acceleration factor is

$$\mathcal{AF}_{\text{BC}}(X_1) = \begin{cases} \left[ \exp \left( \dfrac{X_{1U}^\lambda - X_1^\lambda}{\lambda} \right) \right]^{\gamma_1}, & \text{if } \lambda \neq 0, \\ \left( \dfrac{X_{1U}}{X_1} \right)^{\gamma_1}, & \text{if } \lambda = 0, \end{cases}$$

where $X_{1U}$ are use conditions for the $X_1$ accelerating variable. $\mathcal{AF}_{\text{BC}}(X_1)$ is monotone increasing in $X_1$ if $\gamma_1 < 0$ and monotone decreasing in $X_1$ if $\gamma_1 > 0$.

EXAMPLE 11 (*Spring life test data*). Meeker, Escobar and Zayac (2003) analyze spring accelerated life test data. Time is in units of kilocycles to failure. The explanatory variables are processing temperature (Temp) in degrees Fahrenheit, spring compression displacement (Stroke) in mils, and the categorical variable Method which takes the values New or Old. Springs that had not failed after 5000 kilocycles were coded as "Suspended." At the condition 50 mils, 500 °F and the New processing method, there were no failures before 5000 kilocycles. All of the other conditions had at least some failures, and at five of the twelve conditions all of the springs failed. At some of the conditions, one or more of the springs had not failed after 5000 kilocycles.

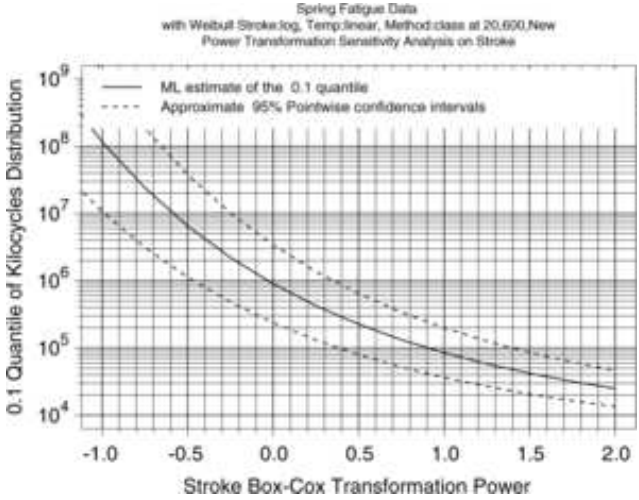

Fig. 12. *Plot of the ML estimate of the 0.10 quantile of spring life at 20 mils, 600 °F, using the new method versus the Stroke displacement Box–Cox transformation parameter λ with 95% confidence limits.*

Figure 12 (see Meeker, Escobar and Zayac, 2003) is a plot of the 0.10 Weibull quantile estimates versus λ from −1 and 2. Approximate confidence intervals are also given. The plot illustrates the sensitivity of the 0.10 quantile estimate to the Box–Cox transformation. Note that λ = 0 corresponds to the log-transformation that is commonly used in fatigue life versus stress models. Also, λ = 1 corresponds to no transformation (or, more precisely, a linear transformation that affects the regression parameter values but not the underlying structure of the model). Figure 12 shows that fatigue life decreases by more than an order of magnitude as λ moves from 0 to 1. In particular, the ML estimate of the 0.10 quantile decreases from 900 megacycles to 84 megacycles when λ is changed from 0 to 1.

Figure 13 is a profile likelihood plot for the Box–Cox λ parameter, providing a visualization of what the data say about the value of this parameter. In this case the peak is at a value of λ close to 0; this is in agreement with the commonly used fatigue life/stress model. Values of λ close to 1 are less plausible, but cannot be ruled out, based on these data alone. The engineers, based on experience with the same failure mode and similar materials, felt that the actual value of λ was near 0 (corresponding to the log-transformation) and almost certainly less than 1. Thus a conservative decision could be made by designing with an assumed value of λ = 1. Even the somewhat optimistic evaluation using λ = 0 would not meet the 500 megacycle target life.

Meeker, Escobar and Zayac (2003) also discuss the sensitivity to the assumed form of the temperature-life relationship and the sensitivity to changes in the assumed distribution.

## 9. ACCELERATION MODELS WITH MORE THAN ONE ACCELERATING VARIABLE

Some accelerated tests use more than one accelerating variable. Such tests might be suggested when it is known that two or more potential accelerating variables contribute to degradation and failure. Using two or more variables may provide needed time-acceleration without requiring levels of the individual accelerating variables to be too high. Some accelerated tests include engineering variables that are not accelerating variables. Examples include material type, design, operator, and so on.

### 9.1 Generalized Eyring Relationship

The generalized Eyring relationship extends the Eyring relationship in Section 5.3, allowing for one or more nonthermal accelerating variables (such as humidity or voltage). For one additional nonthermal accelerating variable $X$, the model, in terms of reaction rate, can be written as

$$
\begin{aligned}
\mathcal{R}(\texttt{temp}, X) \\
(13) \qquad = \gamma_0 \times (\texttt{temp K})^m \times \exp\left(\frac{-\gamma_1}{k \times \texttt{temp K}}\right)
\end{aligned}
$$

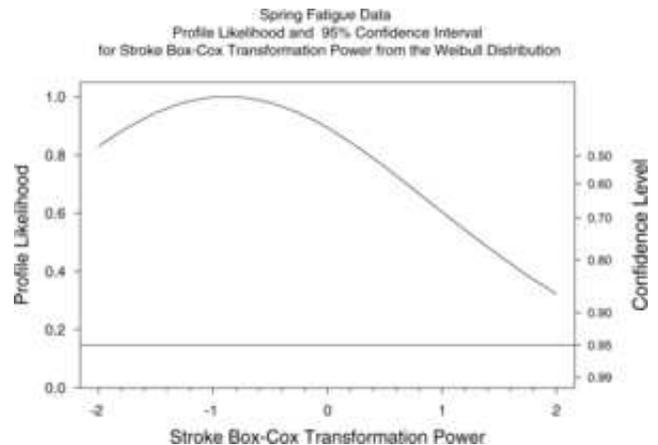

Fig. 13. *Profile likelihood plot for the Stroke Box–Cox transformation parameter λ in the spring life model.*

$$\times \exp\left(\gamma_2 X + \frac{\gamma_3 X}{k \times \texttt{temp}\,\mathrm{K}}\right)$$

where $X$ is a function of the nonthermal stress. The parameters $\gamma_1 = E_a$ (activation energy) and $\gamma_0$, $\gamma_2$, $\gamma_3$ are characteristics of the particular physical/chemical process. Additional factors like the one on the right-hand side of (13) can be added for other nonthermal accelerating variables.

In the following sections, following common practice, we set $(\texttt{temp}\,\mathrm{K})^m = 1$, using what is essentially the Arrhenius temperature-acceleration relationship. These sections describe some important special-case applications of this more general model. If the underlying model relating the degradation process to failure is a SAFT model, then, as in Section 5.2, the generalized Eyring relationship can be used to describe the relationship between times at different sets of conditions $\texttt{temp}$ and $X$. In particular, the acceleration factor relative to use conditions $\texttt{temp}_U$ and $X_U$ is

$$\mathcal{AF}(\texttt{temp}, X) = \frac{\mathcal{R}(\texttt{temp}, X)}{\mathcal{R}(\texttt{temp}_U, X_U)}.$$

The same approach used in Section 5.4 shows the effect of accelerating variables on time to failure. For example, suppose that $T(\texttt{temp}_U)$ (time at use or some other baseline temperature) has a log-location-scale distribution with parameters $\mu_U$ and $\sigma$. Then $T(\texttt{temp})$ has the same log-location-scale distribution with

$$(14) \quad \begin{aligned} \mu &= \mu_U - \log[\mathcal{AF}(\texttt{temp}, X)] \\ &= \beta_0 + \beta_1 x_1 + \beta_2 x_2 + \beta_3 x_1 x_2 \end{aligned}$$

where $\beta_1 = E_a$, $\beta_2 = -\gamma_2$, $\beta_3 = -\gamma_3$, $x_1 = 11605/(\texttt{temp}\,\mathrm{K})$, $x_2 = X$ and $\beta_0 = \mu_U - \beta_1 x_{1U} - \beta_2 x_{2U} - \beta_3 x_{1U} x_{2U}$.

### 9.2 Temperature-Voltage Acceleration Models

Many different models have been used to describe the effect of the combination of temperature and voltage on acceleration. For instance, Meeker and Escobar (1998, Section 17.7) analyzed data from a study relating voltage and temperature to the failure of glass capacitors. They modeled the location parameter of log-lifetime as a simple linear function of $\texttt{temp}\,^\circ\mathrm{C}$ and $\texttt{volt}$. The generalized Eyring relationship in Section 13 can also be used with $X = \log(\texttt{volt})$, as done in Boyko and Gerlach (1989). Klinger (1991) modeled the Boyko and Gerlach (1989) data by including second-order terms for both accelerating variables.

To derive the time-acceleration factor for the extended Arrhenius relationship with $\texttt{temp}$ and $\texttt{volt}$, one can follow steps analogous to those outlined in Section 8.2. Using (13) with $X = \log(\texttt{volt})$, one obtains

$$\begin{aligned} \mathcal{R}&(\texttt{temp}, \texttt{volt}) \\ &= \gamma_0 \times \exp\left(\frac{-E_a}{k \times \texttt{temp}\,\mathrm{K}}\right) \\ &\quad \times \exp\left[\gamma_2 \log(\texttt{volt}) + \frac{\gamma_3 \log(\texttt{volt})}{k \times \texttt{temp}\,\mathrm{K}}\right]. \end{aligned}$$

Again, failure occurs when the dielectric strength crosses the applied voltage stress, that is, $\mathcal{D}(t) = \texttt{volt}$. This occurs at time

$$T(\texttt{temp}, \texttt{volt}) = \frac{1}{\mathcal{R}(\texttt{temp}, \texttt{volt})}\left(\frac{\texttt{volt}}{\delta_0}\right)^{\gamma_1}.$$

From this, one computes

$$\begin{aligned} \mathcal{AF}&(\texttt{temp}, \texttt{volt}) \\ &= \frac{T(\texttt{temp}_U, \texttt{volt}_U)}{T(\texttt{temp}, \texttt{volt})} \\ &= \exp[E_a(x_{1U} - x_1)] \times \left(\frac{\texttt{volt}}{\texttt{volt}_U}\right)^{\gamma_2 - \gamma_1} \\ &\quad \times \{\exp[x_1 \log(\texttt{volt}) - x_{1U} \log(\texttt{volt}_U)]\}^{\gamma_3}, \end{aligned}$$

where $x_{1U} = 11605/(\texttt{temp}_U\,\mathrm{K})$ and $x_1 = 11605/(\texttt{temp}\,\mathrm{K})$. When $\gamma_3 = 0$, there is no interaction between temperature and voltage. In this case, $\mathcal{AF}(\texttt{temp}, \texttt{volt})$ can be factored into two terms, one that involves temperature only and another term that involves voltage only. Thus, if there is no interaction, the contribution of temperature (voltage) to acceleration is the same at all levels of voltage (levels of temperature).

### 9.3 Temperature-Current Density Acceleration Models

d'Heurle and Ho (1978) and Ghate (1982) studied the effect of increased current density ($\mathrm{A/cm^2}$) on electromigration in microelectronic aluminum conductors. High current densities cause atoms to move more rapidly, eventually causing extrusion or voids that lead to component failure. ATs for electromigration often use increased current density and temperature to accelerate the test. An extended Arrhenius relationship could be appropriate for such data.

In particular, when $T$ has a log-location-scale distribution, then (13) applies with $x_1 = 11605/\texttt{temp}\,\mathrm{K}$, $x_2 = \log(\texttt{current})$. The model with $\beta_3 = 0$ (without interaction) is known as "Black's equation" (Black, 1969).

EXAMPLE 12 [*Light emitting diode (LED) reliability*]. A degradation study on a light emitting diode (LED) device was conducted to study the effect of current and temperature on light output over time and to predict life at use conditions. A unit was said to have failed if its light output was reduced to 60% of its initial value. Two levels of current and six levels of temperature were used in the test. Figure 14 shows the LED light output data ver-sus time in hours, in the square-root scale. No units had failed during the test. For a simple method of degradation analysis, predicted pseudo failure times are obtained by using ordinary least squares to fit a line through each sample path on the square-root scale, for "Hours," and the linear scale for "Relative Change." Figure 15 shows the ML fit of the Arrhenius-inverse power relationship lognormal model (with no interaction) for the pseudo failure LED data. The data at 130 °C and 40 mA were omitted in the model fitting because it was determined that a new failure mode had manifested itself at that highest level of test conditions (initial efforts by engineers to use the bad data had resulted in physically impossible estimates of life at the use conditions). Figure 15 also shows the estimate of $F(t)$ at use conditions of 20 °C and 20 mA.

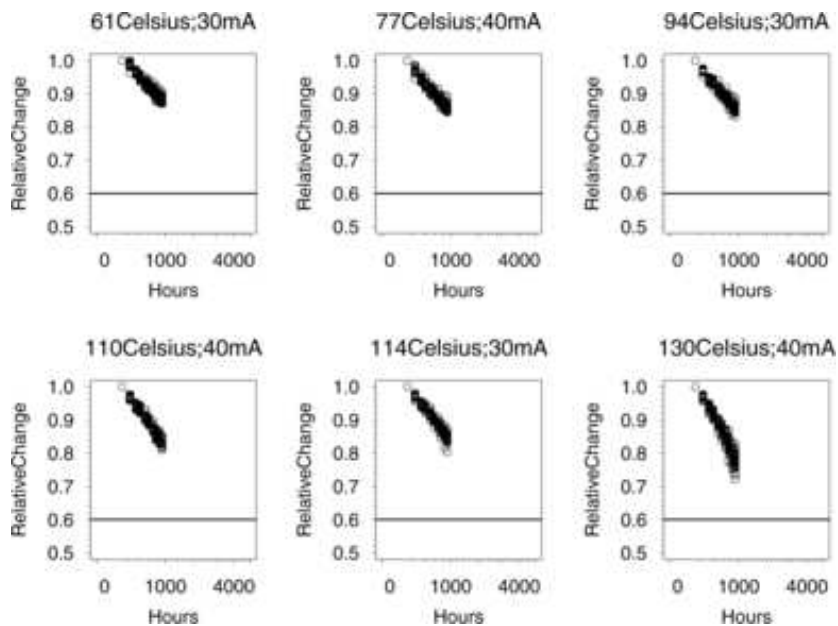

FIG. 14. *Relative change in light output from* 138 *hours at different levels of temperature and current. Relative change is in the linear scale and hours is in the square-root scale which linearizes the response as a function of time.*

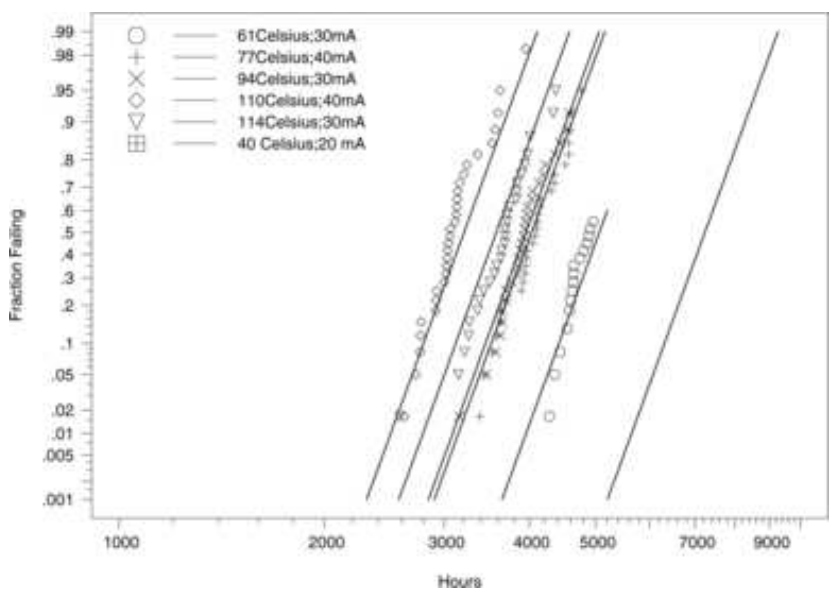

FIG. 15. *LED device data. Lognormal multiple probability plot showing the fitted Arrhenius-inverse power relationship log-normal model (with no interaction) for the failure LED data. The plots also show the estimate of $F(t)$ at use conditions,* 20 °C *and* 20 *mA.*

## 9.4 Temperature-Humidity Acceleration Models

Relative humidity is another environmental variable that can be combined with temperature to accelerate corrosion or other chemical reactions. Examples of applications include paints and coatings, electronic devices and electronic semiconductor parts, circuit boards, permalloy specimens, foods and pharmaceuticals. Although most ALT models that include humidity were derived empirically, some humidity models have a physical basis. For example, Gillen and Mead (1980) and Klinger (1991) studied kinetic models relating aging with humidity. LuValle, Welsher and Mitchell (1986) provided physical basis for studying the effect of humidity, temperature and voltage on the failure of circuit boards. See Boccaletti et al. (1989), Chapter 2 of Nelson (1990), Joyce et al. (1985), Peck (1986) and Peck and Zierdt (1974) for ALT applications involving temperature and humidity.

The extended Arrhenius relationship (13) applied to ALTs with humidity uses $x_1 = 11605/\texttt{temp K}$, $x_2 = \log(\texttt{RH})$ and $x_3 = x_1 x_2$ where RH is a proportion denoting relative humidity. The case when $\beta_3 = 0$ (no temperature-humidity interaction) is known as "Peck's relationship" and was used by Peck (1986) to study failures of epoxy packing. Klinger (1991) suggested the term $x_2 = \log[\texttt{RH}/(1 - \texttt{RH})]$ instead of $\log(\texttt{RH})$. This alternative relationship is based on a kinetic model for corrosion.

## 9.5 Modeling Photodegradation Temperature and Humidity Effects

When modeling photodegradation, as described in Section 7, it is often necessary to account for the effect of temperature and humidity. The Arrhenius rate reaction model (3) can be used to scale time (or dosage) in the usual manner. Humidity is also known to affect photodegradation rate. Sometimes the rate of degradation will be directly affected by moisture content of the degrading material. In this case one can use a model such as described in Burch, Martin and VanLandingham (2002) to predict moisture content as a function of relative humidity.

Combining these model terms with the log of total effective UV dosage from (8) gives

$$\log(d; \mathrm{CF}, p) = \log[\mathrm{D_{Tot}}(t)] + p \times \log(\mathrm{CF}),$$

$$\mu = \beta_0 + \frac{E_a}{k \times \texttt{temp K}} + C \times \mathrm{MC(RH)},$$

where $\texttt{temp K}$ is temperature in kelvin, $\mathrm{MC(RH)}$ is a model prediction of moisture content, as a function of relative humidity, $k$ is Boltzmann's constant, $E_a$ is a quasi-activation energy, and $\beta_0$ and $C$ are parameters that are characteristic of the material and the degradation process. Figure 16 shows the same data displayed in Figure 7, except that the time scale for the data has been adjusted for the differences in both the neutral density filters and the two different levels of temperature, bringing all data to the scale of a 100% neutral density filter and 55 °C.

## 9.6 The Danger of Induced Interactions

As illustrated in Example 13 of Pascual, Meeker and Escobar (2006) (using data related to Example 12 of this paper), interactions can cause difficulty in the interpretation of accelerated test results. Initially, in that example, an interaction term had been used in the model fitting to provide a model that fits the data better at the high levels of temperature and current density that had been used in the test. Extrapolation to the use conditions, however, produced estimates of life that were shorter than the test conditions! The problem was that with the interaction term, there was a saddle point in the response surface, outside of the range of the data. Extrapolation beyond the saddle point resulted in nonsensical predictions that lower temperature and current would lead to shorter life (in effect, the extrapolation was using a quadratic model).

It is important to choose the definition of accelerated test experimental factors with care. Inappropriate choices can induce strong interactions between the factors. Suppose, for example, that there is no interaction between the factors voltage stress and

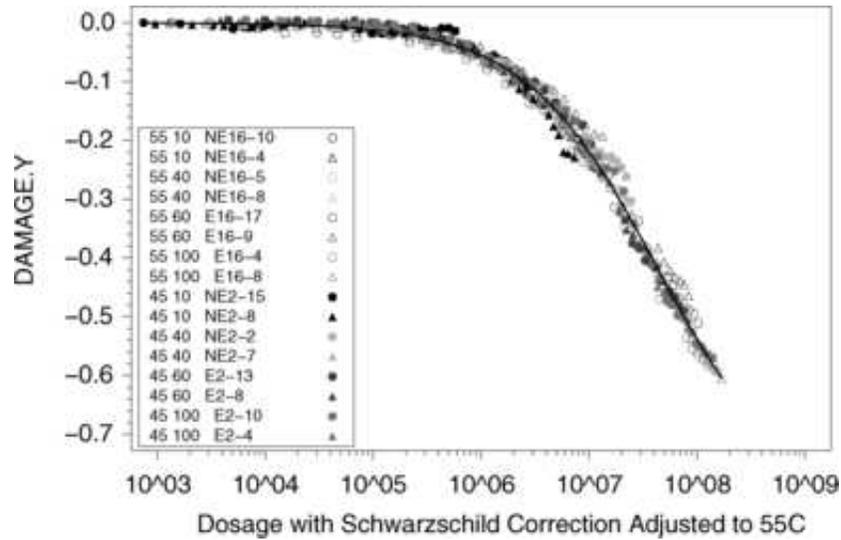

Fig. 16. *Sample paths for wave number* 1510 $cm^{-1}$ *and band-pass filter with nominal center at* 306 *nm for different combinations of temperature (*45 *and* 55*)* °C *and neutral density filter (passing* 10, 40, 60 *and* 100% *of photons across the UV-B spectrum).*

specimen thickness in an acceleration model for dielectric failure. Then

$$\mu = \beta_0 + \beta_1 \text{Thickness} + \beta_2 \text{Voltage Stress}$$

where thickness is measured in mm and Voltage Stress = Voltage/Thickness is measured in V/mm. If the model is written in terms of thickness and voltage,

$$\mu = \beta_0 + \beta_1 \text{Thickness}$$
$$+ \beta_2 \text{Voltage Stress} \times \text{Thickness}.$$

Thus, the variables voltage and thickness would have a strong interaction.

Similarly, if RH and temperature have no interaction in the acceleration model for a corrosion failure mechanism, it is easy to show that the strong effect that temperature has on saturation vapor pressure would imply that the factors temperature and vapor pressure would have a strong interaction.

These concerns are related to the "sliding factor" ideas described in Phadke (1989, Section 6.4).

## 10. COMMENTS ON THE APPLICATION OF ACCELERATION MODELS

### 10.1 Concerns About Extrapolation

All accelerated tests involve extrapolation. The use of extrapolation in applications requires justification. It is always best when the needed justification comes from detailed physical/chemical knowledge of the effect of the accelerating variable on the failure mechanism. As a practical matter, however, such knowledge is often lacking, as are time and resources for acquiring the needed knowledge. Empirical relationships are often used as justification, but rarely are data available to check the relationship over the entire range of interest for the accelerating variable(s).

Evans (1977) makes the important point that the need to make rapid reliability assessments and the fact that accelerated tests may be "the only game in town" are not sufficient to *justify* the use of the method. Justification must be based on physical models or empirical evidence. Evans (1991) describes difficulties with accelerated testing and suggests the use of sensitivity analysis, such as that described in Meeker, Escobar and Zayac (2003). He also comments that acceleration factors of 10 "are not unreasonable" but that "factors much larger than that tend to be figments of the imagination and lots of correct but irrelevant arithmetic."

### 10.2 Some Basic Guidelines

Some guidelines for the use of acceleration models include:

- ATs must generate the same failure mode occurring in the field.

  Generally, accelerated tests are used to obtain information about one particular, relatively simple failure mechanism (or corresponding degradation measure). If there is more than one fail-ure mode, it is possible that the different failure mechanisms will be accelerated at different rates. Then, unless this is accounted for in the modeling and analysis, estimates could be seriously incorrect when extrapolating to lower use levels of the accelerating variables.

- Accelerating variables should be chosen to correspond with variables that cause actual failures.
- It is useful to investigate previous attempts to accelerate failure mechanisms similar to the ones of interest. There are many research reports and papers that have been published in the physics of failure literature. The annual *Proceedings of the International Reliability Physics Symposium*, sponsored by the IEEE Electron Devices Society and the IEEE Reliability Society, contain numerous articles describing physical models for acceleration and failure.
- Accelerated tests should be designed, as much as possible, to minimize the amount of extrapolation required, as described in Chapters 20 and 21 of Meeker and Escobar (1998). High levels of accelerating variables can cause extraneous failure modes that would never occur at use levels of the accelerating variables. If extraneous failures are not recognized and properly handled, they can lead to seriously incorrect conclusions. Also, the relationship may not be accurate enough over a wide range of acceleration.
- In practice, it is difficult or impractical to verify acceleration relationships over the entire range of interest. Of course, accelerated test data should be used to look for departures from the assumed acceleration model. It is important to recognize, however, that the available data will generally provide very little power to detect anything but the most serious model inadequacies. Typically there is no useful diagnostic information about possible model inadequacies at accelerating variable levels close to use conditions.
- Simple models with the right shape have generally proven to be more useful than elaborate multiparameter models.
- Sensitivity analyses should be used to assess the effect of perturbing uncertain inputs (e.g., inputs related to model assumptions).
- Accelerated test programs should be planned and conducted by teams including individuals knowledgeable about the product and its use environment, the physical/chemical/mechanical aspects of the failure mode, and the statistical aspects of the design and analysis of reliability experiments.

## 11. FUTURE RESEARCH IN THE DEVELOPMENT OF ACCELERATED TEST MODELS

Research in the development of accelerated test models is a multidisciplinary activity. Statisticians have an important role to play on the teams of scientists that develop and use accelerated test models. On such a team engineers and scientists are primarily responsible for:

- Identifying and enumerating possible failure modes and, for new products, predicting all possible life-limiting failure modes.
- Understanding the physical/chemical failure mechanisms that lead to a product's failure modes and for identifying accelerating variables that can be used to accelerate the failure mechanism.
- Suggesting deterministic physical/chemical mathematical relationships between the rate of the failure mechanisms and the accelerating variable(s). When such a relationship is not available, they may be able to provide guidance from standard practice or previous experience with similar products and materials.

Statisticians are important for:

- Planning appropriate experiments. Accelerated test programs often start with simple experiments to understand the failure modes that can occur and the behavior of mechanisms that can cause failures. Use of traditional methods for designed experiments is important for these tests. In addition, there may be special features of accelerated tests (e.g., censoring) that require special test planning methods (see Chapter 6 of Nelson, 1990, and Chapter 20 of Meeker and Escobar, 1998, for discussion of accelerated test planning).
- Providing expertise in the analysis of data arising from preliminary studies and the accelerated tests themselves. Features such as censoring, multiple failure modes and models that are nonlinear in the parameters are common. Methods for detecting model departures are particularly important. Model departures may suggest problems with the data, sources of experimental variability that might be eliminated or problems with the suggested model that may suggest changes to the proposed model.
- Identifying sources of variability in experimental data (either degradation data or life data) that reflect actual variability in the failure mechanism.

For this purpose it is generally useful to compare field data. In most applications there will be additional variability in field data and it is important to understand the differences in order to design appropriate laboratory experiments and to be able to draw useful conclusions and predictions from laboratory data.

- Working within cross-disciplinary teams to develop statistical models of failure and acceleration based on fundamental understanding of failure mechanisms. With fundamental understanding of failure mechanisms and knowledge of sources of variability, it is possible to develop, from first principles, failure-time distributions and acceleration models. Better, more credible, models for extrapolation should result from such modeling efforts if the assumptions and other inputs are accurate. Some examples of where such models have been developed include

  - Fisher and Tippett (1928) derived the asymptotic distributions of extreme values and these results provide fundamental motivation for use of distributions such as the Weibull distribution (one of the three distributions of minima) in reliability applications in which failure is caused by the first of many similar possible events (e.g., the failure of a large system with many similar possible failure points, none of which dominates in the failure process).

  - Tweedie (1956), in modeling electrophoretic measurements, used the distribution of first passage time of a Brownian motion with drift to derive the inverse Gaussian distribution.

  - Birnbaum and Saunders (1969), in modeling time to fracture from a fatigue crack growth process, derived a distribution that is today known as the Birnbaum–Saunders distribution. This distribution can be thought of as a discrete-time analog of the inverse Gaussian distribution.

  - Meeker and LuValle (1995), using a kinetic model for the growth of conducting filaments, developed a probability distribution, dependent on the level of relative humidity, to predict the failure time of printed circuit boards.

  - Meeker, Escobar and Lu (1998), using a kinetic model for chemical degradation inside an electronic amplifying device, developed a probability distribution which, when combined with the Arrhenius model, could be used to predict the failure time of the devices.

  - LuValle et al. (1998) and LuValle, Lefevre and Kannan (2004, pages 200–206) used physics-based models to describe degradation processes.

## ACKNOWLEDGMENTS

We want to thank two anonymous referees who provided valuable comments on an earlier version of this paper. These comments led to important improvements in the paper. Charles Annis and Michael LuValle provided us, respectively, helpful descriptions of methods for fatigue testing and modeling chemical processes related to ATs.

Figures 4 and 11 originally appeared in Meeker and Escobar (1998) and Figures 12 and 13 originally appeared in Meeker, Escobar and Zayac (2003) and appear in this paper with permission of the copyright owner, John Wiley and Sons, Inc.